\documentclass[conference, compsoc]{IEEEtran}

\usepackage{graphicx}
\usepackage{float}
\usepackage{amsmath}
\usepackage{amssymb}
\usepackage{tikz}
\usepackage[binary-units=true, group-separator={,}, group-minimum-digits={3}]{siunitx}
\usetikzlibrary{shapes,arrows,positioning,calc}
\usepackage{cite}
\usepackage{arydshln}
\usepackage{xspace}
\usepackage{xurl}
\usepackage{algorithm}
\usepackage[noend]{algpseudocode}

\usepackage{wasysym}
\usepackage{marvosym}
\usepackage{tabularx}
\usepackage{xcolor}
\usepackage{fp}

\usepackage{booktabs}
\usepackage{tabularx}
\usepackage{multirow}
\usepackage{multicol}
\usepackage{colortbl}
\usepackage{blindtext}
\usepackage{tablefootnote}
\usepackage{threeparttable}

\usepackage{pifont}
\usepackage{paralist}

\usepackage[normalem]{ulem}
\usepackage{color}
\usepackage{subcaption}

\usepackage{cuted}
\usepackage{caption}

\definecolor{deepgreen}{RGB}{34,139,34}
\definecolor{deepred}{RGB}{204,51,51}
\newcommand{\up}[1]{\textcolor{deepgreen}{$\blacktriangle$\,#1}}
\newcommand{\down}[1]{\textcolor{deepred}{$\blacktriangledown$\,#1}}

\definecolor{myblue}{HTML}{87CEEB}
\definecolor{darktangerine}{rgb}{1.0, 0.66, 0.07}
\definecolor{darkgreen}{rgb}{0.31, 0.78, 0.47}
\usepackage[]{hyperref}
\hypersetup{
    colorlinks = true,
    linkcolor = myblue,
    anchorcolor = darkgreen,
    citecolor = darkgreen,
    filecolor = darkgreen,
    urlcolor = myblue
}
\usepackage{minibox}
\usepackage{tcolorbox}
\newtcolorbox{mybox}[2][]{%
  colback      = gray!5!white,
  colframe     = gray!75!black,
  fonttitle    = \bfseries,
  colbacktitle =  gray!85!black,
  title        = #2,#1,
}

\newlength\maxlen
\newcommand\appbar[3][darkgreen!30]{%
  \FPeval\result{round(#3/#2:4)}%
  \rlap{\textcolor{#1}{\hspace*{\dimexpr-\tabcolsep+3\arrayrulewidth}%
        \rule[-.5\ht\strutbox]{\result\maxlen}{1.5\ht\strutbox}}}%
  \makebox[\dimexpr\maxlen-0.2\tabcolsep+\arrayrulewidth][r]{\SI{#3}{}}}

\def\header{NumRatings}
\newlength\maxlenweb
\newcommand\webbar[3][darkgreen!30]{%
  \FPeval\result{round(#3/#2:4)}%
  \rlap{\textcolor{#1}{\hspace*{\dimexpr-\tabcolsep+3\arrayrulewidth}%
        \rule[-.5\ht\strutbox]{\result\maxlenweb}{1.5\ht\strutbox}}}%
  \makebox[\dimexpr\maxlenweb-0.2\tabcolsep+\arrayrulewidth][r]{\SI{#3}{}}}

\def\headerweb{MonthlyVisit}
\newcolumntype{?}{!{\vrule width 1pt}}

\newcounter{note}[section]

\colorlet{Mycolor1}{green!10!orange!90!}

\newcommand{\appref}[1]{\hyperref[#1]{Appendix~\ref{#1}}}

\newcommand{\etc}{\textit{etc.}\xspace}
\newcommand{\eg}{\textit{e.g.}\xspace}
\newcommand{\ie}{\textit{i.e.}\xspace}

\newcommand{\aka}{\textit{a.k.a.}\xspace}
\newcommand{\etal}{\textit{et al.}\xspace}

\newcommand{\minor}[1]{\textcolor{black}{#1}}
\newcommand{\minortwo}[1]{\textcolor{black}{#1}}

\newcommand{\cmark}{\ding{51}}%
\newcommand{\xmark}{\ding{55}}%

\newcounter{packednmbr}

\newenvironment{packeditemize}{
\begin{list}{$\bullet$}{
\setlength{\labelwidth}{5pt}
\setlength{\itemsep}{3pt}
\setlength{\leftmargin}{\labelwidth}
\addtolength{\leftmargin}{\labelsep}
\setlength{\parindent}{0pt}
\setlength{\listparindent}{\parindent}
\setlength{\parsep}{1pt}
\setlength{\topsep}{1pt}}}{\end{list}}

\renewcommand{\paragraph}[1]{\vspace{0.02in}\noindent{\bf{#1}.}}
\newcommand{\ignore}[1]{}

\newcommand{\sysname}{\textsc{BraVeSpy}\xspace}

\newcommand*\myYescirc{\CIRCLE\xspace}

\newcommand*\myNocirc{\Circle\xspace}

\begin{document}



%





\title{When VR Meets BCI: (Un)Observable Brainwave-aware Privacy Reconstruction in the Metaverse via Unrestricted Inbuilt Motion Sensors}





%
\author{\IEEEauthorblockN{Tao Ni\IEEEauthorrefmark{1}\quad
Zehua Sun\IEEEauthorrefmark{2}\quad
Qingchuan Zhao\IEEEauthorrefmark{3}\textsuperscript{(\Letter)}\quad
Wei-Bin Lee\IEEEauthorrefmark{4}\IEEEauthorrefmark{5}\quad
Cong Wang\IEEEauthorrefmark{3}\textsuperscript{(\Letter)}}
\IEEEauthorblockA{\IEEEauthorrefmark{1}King Abdullah University of Science and Technology \quad \IEEEauthorrefmark{2}National University of Singapore\quad\\ \IEEEauthorrefmark{3}City University of Hong Kong\quad \IEEEauthorrefmark{4}Feng Chia University\quad\\ \IEEEauthorrefmark{5}Information Security Research Center, Hon Hai Research Institute}}

\maketitle







%


\begin{abstract}

Metaverse devices, such as virtual reality (VR), have seen substantial development and widespread applications in numerous areas.
Although recent studies have revealed privacy leakages in VR, these vulnerabilities were limited in the scope of observable behaviors in virtual scenes (\eg, \textit{what a user is seeing}).
In this work, we uncover the feasibility of going beyond the scope of observable user behaviors to unobservable \minortwo{brain EEG-correlated representations} (\eg, \textit{what a user is \minortwo{perceiving}}) by leveraging unrestricted motion sensors in VR headsets to reconstruct brain EEG signals, a seemingly neglected but promising vector.
The insight is that the inbuilt motion sensors (\eg, accelerometers) in the VR headset can capture subtle vibrations induced by pupillary responses, which are highly correlated with users' visual stimuli and in-brain perceptions.

Therefore, we design and implement \sysname to systematically investigate and demonstrate the feasibility of this severe privacy leakage originating from
brain \minortwo{EEG-correlated representations} reconstructed from variations of inbuilt motion sensors.
Our extensive evaluation results from different VR devices show that \sysname, for the first time in the Metaverse, can reveal unobservable privacy, where we successfully unveiled perceptive images in the brain with $52.0\%$--$67.2\%$ accuracy.
In particular, we also find that 
\sysname outperforms the current approaches that are limited to coarse-grained inference of observable behaviors and achieves over $85.0\%$ accuracy in inferring user activity-related sensitive information, such as fingerprinting websites, apps, and streaming videos, and over $96.0\%$ accuracy in user de-anonymization, gaze movement tracking, and virtual keystroke inference.
\looseness=-1

\end{abstract}

\section{Introduction}
\label{sec:introduction}

Recent years have witnessed exponential advancements in Metaverse technology, such as virtual reality (VR), which aims to create a simulated environment with lifelike scenes and objects to
give users the sensation of being fully immersed in their surroundings
through a VR headset.
Typically, it consists of a head-mounted display (HMD) and inbuilt sensors that allow VR applications to measure user's motions and postures and to support real-time interactions by adjusting the visual content accordingly.
Thus, massive sensor data access has become inevitable in VR apps, and most existing operating systems enforce limited constraints on this data access.
For example, according to a recent empirical study, almost $42\%$ VR apps from the Meta Oculus Quest store~\cite{oculusapps} collect motion sensor data and have no permission requests in the manifest files of the apps~\cite{guo2024empirical}.

Unfortunately, the inevitable access and collection of sensor data in VR applications is expected to result in severe leakage of observable user behaviors, such as keystrokes~\cite{aviv2012practicality, simon2013pin, lu2019keylistener}, the launch of websites/apps~\cite{matyunin2019magneticspy, pan2021magthief, ning2018deepmag}, and speeches~\cite{anand2018speechless, ba2020learning, hu2022accear}, because a large body of research on the smartphone platform has well studied and unveiled the vulnerabilities through which sensor data can leak such sensitive privacy.
Following that line of research, recent security studies specific to the VR platform have also found that its sensor data could reveal observable and measurable privacy issues, including virtual keyboard inputs (\eg, \cite{luo2022holologger, slocum2023going, wu2023privacy, zhang2023s, meteriz2022keylogging, lee2025eyes, khalili2026xr}), $360^{\circ}$ streaming videos~\cite{nguyen2024penetration}, user avatar~\cite{meng2023anonymization, nair2023unique, yang2024can}, speech~\cite{shi2021face, cayir2025speak}, and further reconstructing user biometric information, \ie, body fat ratio~\cite{zhang2023passive}, respiration rates and heartbeats~\cite{zhang2023facereader, zhang2025harnessing} or even blood pressure~\cite{ye2025bpsniff}.
Indeed, these findings have successfully complemented the understanding of the attack surface in the VR platform alongside other attack vectors reported in VR, such as GPU profiles~\cite{son2025side}, network traffic~\cite{al2021vr, su2024remote, khalili2026xr}, camera-recorded videos~\cite{gopal2023hidden, meteriz2022keylogging, nguyen2024penetration, khalili2024virtual}, and controllers' button-clicking sounds~\cite{luo2024eavesdropping} or leaked infrared signals~\cite{ni2024non}.
However, these recent security studies on the VR platform have not fully investigated the unique features of the VR platform and are thus limited in the scope of observable privacy leakages.
\looseness=-1

\begin{figure}[t]
    \centering
    \includegraphics[width=\linewidth]{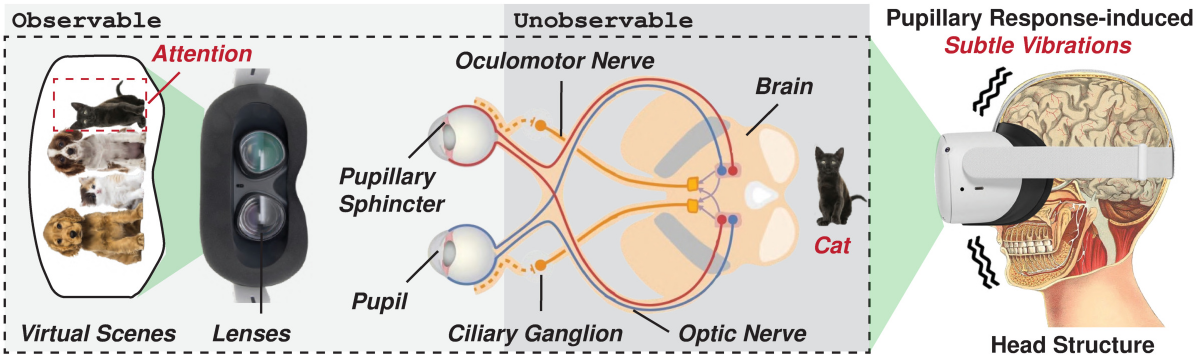}
    \vspace{-0.15in}
    \caption{An illustration of brain perception in virtual scenes.}
    \vspace{-0.25in}
    \label{fig:background_pupillary_response}
\end{figure}

In this work, we found a unique but not fully explored feature of sensors in VR headsets, which are positioned much closer to a user's brain, is promising not only for further exploring much higher granularity of \textit{what a user is seeing or doing} but also in going beyond this scope to \textit{what a user is \minortwo{perceiving}}, \aka, observable behaviors to unobservable behaviors.
This transition is believed to be critical to fully understanding the potential of user privacy leakage in VR due to the inconsistency between user visual perception and brain perception~\cite{smythies2005brain, dror2005perception}, which is a
well-established theoretical and empirical foundation in the field of brain science and cognitive neuroscience.
When it comes to the VR platform, it is noted that brain perception is selective as the brain constantly decides what information from the HMD is important enough to reach our consciousness. 
Thus, a large part of the sensory information from virtual scenes that constantly arrives through our senses is never consciously processed. 
Complex mechanisms in the brain filter the incoming sensory information and shape the representation of the world in human minds~\cite{dwarakanath2023bistability, abbas2024neuroarchitecture}.
For example, \autoref{fig:background_pupillary_response} shows that when the VR user views a picture with multiple objects (\eg, different cats and dogs) in the virtual scene, his/her brain selectively processes only the object of attention (\eg, black domestic shorthair cat)~\cite{lockhofen2021neurochemistry}.
\looseness=-1

Our proposed hypothesis to leverage VR sensors to reveal users' unobservable and fine-grained observable behaviors is believed to be plausible and reasonable
because previous theories have validated that (1) unobservable behaviors (\eg, brain image perceptions) \minortwo{reflected} in the eye responses~\cite{ko2020eyeblink, shahbakhti2021simultaneous, gusso2022more} and brain \minortwo{EEG-correlated reactions}~\cite{bai2023dreamdiffusion, lan2023seeing, davis2022brain},
(2) brain EEG variations or eye responses, in turn, control the user's body reactions and movements~\cite{reis2014methodological, land2006eye}, and (3) these two signals are concentrated in the brain's forehead but are too weak to be measured as the distance increases. However, the VR headset is tightly mounted on the forehead scalp around the eyes, ensuring close proximity to measure high-quality data.
As such, the main challenge lies in the feasibility of bridging the relationship between VR sensors and these two pieces of information.
Our key insight is the observation that the pupillary responses of the human eye under different visual stimuli are correlated with the dilation and contraction of the pupillary sphincter~\cite{slanzi2017combining, park2018infrared} that leads to \textit{subtle facial vibrations} around the eyes.
In particular, these induced subtle vibrations are related to brain EEG signals~\cite{ko2020eyeblink, shahbakhti2021simultaneous, gusso2022more} and could propagate through the VR headset to vibrate inbuilt motion sensors.
Hence, data fluctuations of these motion sensors could reflect pupillary responses and in-depth \minortwo{brain EEG-correlated} variations.
\looseness=-1

\begin{figure}[t]
    \centering
    \includegraphics[width=.95\linewidth]{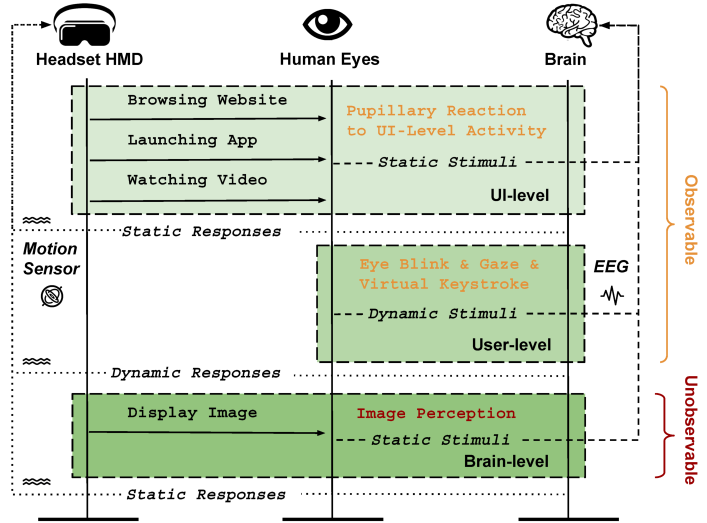}
    \vspace{-0.05in}
    \caption{Three levels of observable (UI-level and user-level) behaviors and unobservable (brain-level) perceptions in VR.}
    \vspace{-0.25in}
    \label{fig:leakage_model}
\end{figure}

Therefore, we design and implement an attack system, \sysname, to systematically explore the potential privacy leakages resulting from the unrestricted motion sensors built into the VR headset. \minortwo{By exploiting the statistical correlation between pupillary response-induced vibrations and brain-level visual perception, \sysname reconstructs EEG-correlated representations from motion sensor data to infer both observable and unobservable user activities.} As depicted in \autoref{fig:leakage_model}, we propose a privacy leakage model that captures the correlations between visual stimuli and pupillary responses at three levels:
\looseness=-1

\begin{packeditemize}
    \item \textbf{Observable UI-level Privacy:} This privacy refers to the content displayed on the HMD. Related works have only studied single coarse-grained privacy, such as keystrokes~\cite{luo2022holologger, slocum2023going, wu2023privacy, luo2024eavesdropping, wang2024gazeploit, su2024remote, meteriz2022keylogging, ni2024non, li2024dangers, gopal2023hidden, al2021vr, lee2025eyes, khalili2026xr}, apps~\cite{son2025side} and $360^{\circ}$ videos~\cite{nguyen2024penetration}. We comprehensively explore more fine-grained UI-level privacy, including fingerprinting websites, and in-app activities such as streaming videos that have not yet been attempted.
    
    \item \textbf{Observable User-level Privacy:} This privacy concern is mainly related to the biometric features of VR users who wear the HMD. Previous works have focused only on user profiles, such as users' identity and gender~\cite{shi2021face, zhang2024safari, nair2023unique, meng2023anonymization, zhu2023soundlock, cayir2025speak}, vital signals such as respiration and heartbeat rates~\cite{zhang2023facereader, zhang2023passive, zhang2025harnessing}, and blood pressures~\cite{ye2025bpsniff}. Our research goes further by exploring user de-anonymization, gaze movements and virtual keystroke inference.
    
    \item \textbf{Unobservable Brain-level Privacy:} This privacy is relevant to the information the brain selects and reaches the user's consciousness, such as brain image perceptions~\cite{bai2023dreamdiffusion, lan2023seeing, davis2022brain}. To our knowledge, \sysname becomes the \textit{first} attempt to reveal such unobservable brain-level perceptions using the reconstructed \minortwo{EEG-correlated representations} from the VR headset's motion sensors.
\end{packeditemize}

We have evaluated \sysname on four popular VR headsets: Meta Oculus Quest 2, Meta Oculus Quest, PICO 4 All-in-One, and HTC Vive Pro. Our evaluation results show that \sysname
can effectively infer sensitive user information
at the aforementioned three levels, and also presents high robustness and transferability when considering conditions with varying impact factors.
In addition, we introduce effective countermeasures to prevent privacy leakage from inbuilt unrestricted motion sensors.

\paragraph{Contributions} We summarize our contributions as follows:

\begin{packeditemize}
    \item We identify a novel attack vector in which adversaries can \minortwo{infer EEG-correlated perceptions} using unrestricted motion sensors in VR headsets, which poses significant threats to user privacy within the emerging Metaverse.
    \item We systematically investigate the variations of \minortwo{brain EEG patterns} with pupillary responses under visual stimuli corresponding to user activities in virtual scenes. Furthermore, we model three levels of observable and unobservable privacy with reconstructed \minortwo{EEG-correlated representations} from visual presentation to brain perceptions.
    \item We design and implement an end-to-end system \sysname to validate the feasibility and practicality of the proposed attack vector, which could not only infer the privacy of user level (\eg, browsing websites, running VR apps, playing streaming video) and user level (\eg, identity, gaze, keystroke), but also in-brain image perceptions, which outperform all previous related studies.
\end{packeditemize}

\section{Preliminary}
\label{sec:background}

\subsection{Unrestricted Motion Sensors in VR Headset}
\label{subsec:background_motion_sensors_vr_headset}

In order to enable immersive interactions in the virtual environment, VR headsets (\eg, Meta Oculus Quest 2) integrate with various embedded sensors such as motion sensors (\eg, accelerometer, gyroscope), which are critical for estimating the orientation of the headset.
However, mainstream VR SDKs and APIs, including OpenXR~\cite{picoopenxrsdk}, Oculus Mobile SDK~\cite{oculusmobilesdk}, and WebXR Device API~\cite{webxrapi}, do not impose restrictions on accessing the motion sensor readings inside the VR headset.
This openness has enabled research, as evidenced by recent studies (\eg, ~\cite{wu2023privacy, zhang2023s, slocum2023going, luo2022holologger, meteriz2022keylogging, ling2019know, ye2025bpsniff, zhang2023facereader, shi2021face, zhang2023passive, cayir2025speak, lee2025eyes}), which demonstrate that adversaries can potentially deploy malicious apps or entice users into accessing malicious web pages, allowing stealthy and continuous logging of VR sensor data in the background without the user's permission.
The data from these unrestricted motion sensors contain a substantial amount of sensitive information that could be maliciously exploited to violate sensitive user privacy.
\looseness=-1

Due to head-mounted characteristics, the unrestricted motion sensors of VR headsets can detect subtle facial vibrations, revealing sensitive biometric data such as gender and identity (\eg, \cite{shi2021face, nair2022exploring, zhang2023passive}) as well as vital signs such as respiration rate, heartbeat~\cite{zhang2023facereader}, and blood pressure~\cite{ye2025bpsniff}.
Furthermore, because these headsets are positioned around the eyes, they can capture subtle facial vibrations triggered by pupillary responses to visual stimuli.
These subtle vibrations around the eyes are linked to forehead EEG signals (\eg, \cite{ko2020eyeblink, shahbakhti2021simultaneous, gusso2022more}), which reflect brain activity. This capability underscores the potential for VR technology to provide insight into a user's physiological and neurological state.
\looseness=-1

\subsection{Pupillary Responses under Visual Stimuli}
\label{subsec:background_pupillary}

Due to the head-mounted design of HMD headsets, the users' eyes become the primary interface for receiving information within the virtual environment and transmitting it to the brain. 
Specifically, eyes visualize virtual scenes by receiving and processing different visual stimuli from various user interfaces and content displays (\aka, pupillary light reflex~\cite{kardon1995pupillary, nozaki2019pupillary}), and the responded visual stimuli are then transmitted to the brain via the optic nerves. 
Furthermore, as the pupils adjust to focus on the displayed content, they present information that varies with their gaze.
These pupil responses and movements influence brain perceptions, which results in different brainwave signals and is widely confirmed by previous studies~\cite{zhang2023recognition, park2018infrared, slanzi2017combining}.
In addition, pupillary responses also induce muscle movements (\eg, dilation and contraction) that cause subtle vibrations, which can be detected by the inbuilt unrestricted motion sensors of the HMD headset that is tightly mounted on the forehead scalp.
Therefore, pupillary responses bridge the correlations between brainwave variations and subtle vibrations on the motion sensors embedded in commercial VR headsets.

\begin{table}[t]
\centering
\begin{threeparttable}
\caption{Comparison with prior studies relevant to VR privacy leakage at two \textit{observable} levels: (\textbf{L1}) \underline{UI-level} visual presentation inference (\raisebox{-0.01in}{\includegraphics[width=9pt]{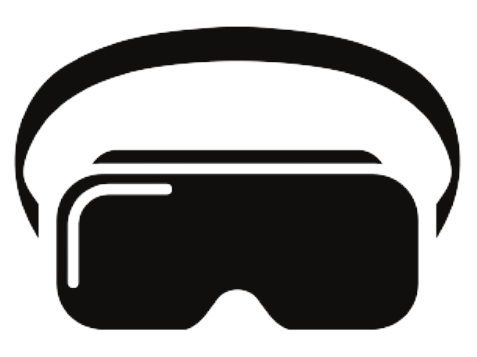}} $\rightarrow$ \raisebox{-0.01in}{\includegraphics[width=8pt]{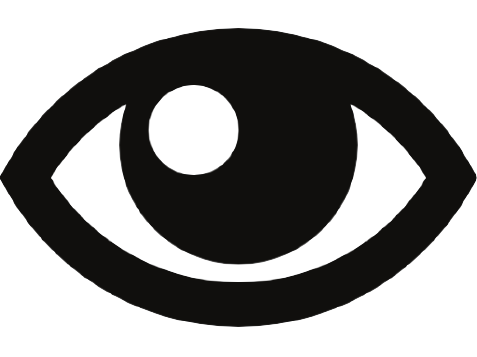}}); (\textbf{L2}) \underline{User-level} identification and gaze recognition (\raisebox{-0.01in}{\includegraphics[width=8pt]{figures/eye_icon.pdf}} $\rightarrow$ \raisebox{-0.01in}{\includegraphics[width=8pt]{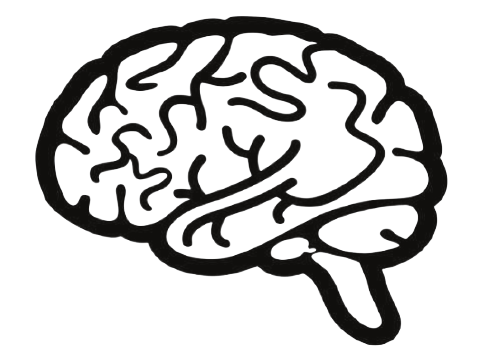}}), and one \textit{unobservable} level: (\textbf{L3}) \underline{Brain-level} \minor{perception inference} (\raisebox{-0.01in}{\includegraphics[width=9pt]{figures/headset_icon.pdf}} $\rightarrow$ \raisebox{-0.01in}{\includegraphics[width=8pt]{figures/eye_icon.pdf}} $\rightarrow$ \raisebox{-0.01in}{\includegraphics[width=8pt]{figures/brain_icon.pdf}}).}
\label{tab:comparison_vr_attacks}
\scriptsize
\setlength{\tabcolsep}{1.5pt}
\renewcommand{\arraystretch}{1.0}
\begin{tabular}{lcccccccc}
\toprule
\textbf{VR Attack} & \textbf{Venue} & \textbf{Attack Vector} & \textbf{M1}  & \textbf{M2} & \textbf{M3} & \textbf{L1} & \textbf{L2} & \textbf{L3} \\ \midrule
\multicolumn{1}{l!{\vrule width 0.5pt}}{VR-Spy~\cite{al2021vr}} & \multicolumn{1}{c!{\vrule width 0.5pt}}{IEEE VR'21} & \multicolumn{1}{c!{\vrule width 0.5pt}}{Wi-Fi CSI} & \multicolumn{1}{c!{\vrule width 0.5pt}}{\xmark} & \multicolumn{1}{c!{\vrule width 0.5pt}}{\cmark} & \multicolumn{1}{c!{\vrule width 0.5pt}}{--} & \multicolumn{1}{c!{\vrule width 0.5pt}}{\myYescirc} & \multicolumn{1}{c!{\vrule width 0.5pt}}{\myNocirc} & \myNocirc \\
\multicolumn{1}{l!{\vrule width 0.5pt}}{Holologger~\cite{luo2022holologger}} & \multicolumn{1}{c!{\vrule width 0.5pt}}{IEEE VR'22} & \multicolumn{1}{c!{\vrule width 0.5pt}}{Motion sensor} & \multicolumn{1}{c!{\vrule width 0.5pt}}{\cmark} & \multicolumn{1}{c!{\vrule width 0.5pt}}{\xmark} & \multicolumn{1}{c!{\vrule width 0.5pt}}{\cmark} & \multicolumn{1}{c!{\vrule width 0.5pt}}{\myYescirc} & \multicolumn{1}{c!{\vrule width 0.5pt}}{\myNocirc} & \myNocirc \\
\multicolumn{1}{l!{\vrule width 0.5pt}}{TyPose~\cite{slocum2023going}} & \multicolumn{1}{c!{\vrule width 0.5pt}}{Security'23} & \multicolumn{1}{c!{\vrule width 0.5pt}}{Motion sensor} & \multicolumn{1}{c!{\vrule width 0.5pt}}{\cmark} & \multicolumn{1}{c!{\vrule width 0.5pt}}{\xmark} & \multicolumn{1}{c!{\vrule width 0.5pt}}{\cmark} & \multicolumn{1}{c!{\vrule width 0.5pt}}{\myYescirc} & \multicolumn{1}{c!{\vrule width 0.5pt}}{\myNocirc} & \myNocirc \\ 
\multicolumn{1}{l!{\vrule width 0.5pt}}{Wu~\etal~\cite{wu2023privacy}} & \multicolumn{1}{c!{\vrule width 0.5pt}}{S\&P'23} & \multicolumn{1}{c!{\vrule width 0.5pt}}{Motion sensor} & \multicolumn{1}{c!{\vrule width 0.5pt}}{\cmark} & \multicolumn{1}{c!{\vrule width 0.5pt}}{\xmark} & \multicolumn{1}{c!{\vrule width 0.5pt}}{\cmark} & \multicolumn{1}{c!{\vrule width 0.5pt}}{\myYescirc} & \multicolumn{1}{c!{\vrule width 0.5pt}}{\myNocirc} & \myNocirc \\
\multicolumn{1}{l!{\vrule width 0.5pt}}{SnoopFinger~\cite{lee2025eyes}} & \multicolumn{1}{c!{\vrule width 0.5pt}}{S\&P'25} & \multicolumn{1}{c!{\vrule width 0.5pt}}{Motion sensor} & \multicolumn{1}{c!{\vrule width 0.5pt}}{\cmark} & \multicolumn{1}{c!{\vrule width 0.5pt}}{\xmark} & \multicolumn{1}{c!{\vrule width 0.5pt}}{\cmark} & \multicolumn{1}{c!{\vrule width 0.5pt}}{\myYescirc} & \multicolumn{1}{c!{\vrule width 0.5pt}}{\myNocirc} & \myNocirc \\
\multicolumn{1}{l!{\vrule width 0.5pt}}{ImmerSpy~\cite{cayir2025speak}} & \multicolumn{1}{c!{\vrule width 0.5pt}}{NDSS'25} & \multicolumn{1}{c!{\vrule width 0.5pt}}{Motion sensor} & \multicolumn{1}{c!{\vrule width 0.5pt}}{\cmark} & \multicolumn{1}{c!{\vrule width 0.5pt}}{\xmark} & \multicolumn{1}{c!{\vrule width 0.5pt}}{\cmark} & \multicolumn{1}{c!{\vrule width 0.5pt}}{\myNocirc} & \multicolumn{1}{c!{\vrule width 0.5pt}}{\myYescirc} & \myNocirc \\ 
\multicolumn{1}{l!{\vrule width 0.5pt}}{Gopal~\etal~\cite{gopal2023hidden}} & \multicolumn{1}{c!{\vrule width 0.5pt}}{Security'23} & \multicolumn{1}{c!{\vrule width 0.5pt}}{Recorded video} & \multicolumn{1}{c!{\vrule width 0.5pt}}{\xmark} & \multicolumn{1}{c!{\vrule width 0.5pt}}{\cmark} & \multicolumn{1}{c!{\vrule width 0.5pt}}{--} & \multicolumn{1}{c!{\vrule width 0.5pt}}{\myYescirc} & \multicolumn{1}{c!{\vrule width 0.5pt}}{\myNocirc} & \myNocirc \\
\multicolumn{1}{l!{\vrule width 0.5pt}}{Intrude~\cite{nguyen2024penetration}} & \multicolumn{1}{c!{\vrule width 0.5pt}}{Security'24} & \multicolumn{1}{c!{\vrule width 0.5pt}}{Recorded video} & \multicolumn{1}{c!{\vrule width 0.5pt}}{\xmark} & \multicolumn{1}{c!{\vrule width 0.5pt}}{\cmark} & \multicolumn{1}{c!{\vrule width 0.5pt}}{--} & \multicolumn{1}{c!{\vrule width 0.5pt}}{\myYescirc} & \multicolumn{1}{c!{\vrule width 0.5pt}}{\myNocirc} & \myNocirc \\
\multicolumn{1}{l!{\vrule width 0.5pt}}{LineTalker~\cite{li2024dangers}} & \multicolumn{1}{c!{\vrule width 0.5pt}}{TIFS'24} & \multicolumn{1}{c!{\vrule width 0.5pt}}{Charging current} & \multicolumn{1}{c!{\vrule width 0.5pt}}{\xmark} & \multicolumn{1}{c!{\vrule width 0.5pt}}{\cmark} & \multicolumn{1}{c!{\vrule width 0.5pt}}{--} & \multicolumn{1}{c!{\vrule width 0.5pt}}{\myYescirc} & \multicolumn{1}{c!{\vrule width 0.5pt}}{\myNocirc} & \myNocirc \\
\multicolumn{1}{l!{\vrule width 0.5pt}}{Heimdall~\cite{luo2024eavesdropping}} & \multicolumn{1}{c!{\vrule width 0.5pt}}{NDSS'24} & \multicolumn{1}{c!{\vrule width 0.5pt}}{Controllers' sound} & \multicolumn{1}{c!{\vrule width 0.5pt}}{\xmark} & \multicolumn{1}{c!{\vrule width 0.5pt}}{\cmark} & \multicolumn{1}{c!{\vrule width 0.5pt}}{--} & \multicolumn{1}{c!{\vrule width 0.5pt}}{\myYescirc} & \multicolumn{1}{c!{\vrule width 0.5pt}}{\myNocirc} & \myNocirc \\
\multicolumn{1}{l!{\vrule width 0.5pt}}{Su~\etal~\cite{su2024remote}} & \multicolumn{1}{c!{\vrule width 0.5pt}}{Security'24} & \multicolumn{1}{c!{\vrule width 0.5pt}}{Unencrypted network} & \multicolumn{1}{c!{\vrule width 0.5pt}}{\xmark} & \multicolumn{1}{c!{\vrule width 0.5pt}}{\cmark} & \multicolumn{1}{c!{\vrule width 0.5pt}}{--} & \multicolumn{1}{c!{\vrule width 0.5pt}}{\myYescirc} & \multicolumn{1}{c!{\vrule width 0.5pt}}{\myNocirc} & \myNocirc \\
\multicolumn{1}{l!{\vrule width 0.5pt}}{VRecKey~\cite{ni2024non}} & \multicolumn{1}{c!{\vrule width 0.5pt}}{NDSS'25} & \multicolumn{1}{c!{\vrule width 0.5pt}}{Infrared signals} & \multicolumn{1}{c!{\vrule width 0.5pt}}{\xmark} & \multicolumn{1}{c!{\vrule width 0.5pt}}{\cmark} & \multicolumn{1}{c!{\vrule width 0.5pt}}{--} & \multicolumn{1}{c!{\vrule width 0.5pt}}{\myYescirc} & \multicolumn{1}{c!{\vrule width 0.5pt}}{\myNocirc} & \myNocirc \\
\multicolumn{1}{l!{\vrule width 0.5pt}}{TwiST~\cite{khalili2026xr}} & \multicolumn{1}{c!{\vrule width 0.5pt}}{NDSS'26} & \multicolumn{1}{c!{\vrule width 0.5pt}}{Wi-Fi packets} & \multicolumn{1}{c!{\vrule width 0.5pt}}{\xmark} & \multicolumn{1}{c!{\vrule width 0.5pt}}{\cmark} & \multicolumn{1}{c!{\vrule width 0.5pt}}{--} & \multicolumn{1}{c!{\vrule width 0.5pt}}{\myYescirc} & \multicolumn{1}{c!{\vrule width 0.5pt}}{\myNocirc} & \myNocirc \\
\multicolumn{1}{l!{\vrule width 0.5pt}}{Zhang~\etal~\cite{zhang2023s}} & \multicolumn{1}{c!{\vrule width 0.5pt}}{Security'23} & \multicolumn{1}{c!{\vrule width 0.5pt}}{Sensor\&System load} & \multicolumn{1}{c!{\vrule width 0.5pt}}{\cmark} & \multicolumn{1}{c!{\vrule width 0.5pt}}{\xmark} & \multicolumn{1}{c!{\vrule width 0.5pt}}{\xmark} & \multicolumn{1}{c!{\vrule width 0.5pt}}{\myYescirc} & \multicolumn{1}{c!{\vrule width 0.5pt}}{\myNocirc} & \myNocirc \\ 
\multicolumn{1}{l!{\vrule width 0.5pt}}{OVRWatcher~\cite{son2025side}} & \multicolumn{1}{c!{\vrule width 0.5pt}}{NDSS'26} & \multicolumn{1}{c!{\vrule width 0.5pt}}{GPU profile} & \multicolumn{1}{c!{\vrule width 0.5pt}}{\cmark} & \multicolumn{1}{c!{\vrule width 0.5pt}}{\xmark} & \multicolumn{1}{c!{\vrule width 0.5pt}}{\xmark} & \multicolumn{1}{c!{\vrule width 0.5pt}}{\myYescirc} & \multicolumn{1}{c!{\vrule width 0.5pt}}{\myNocirc} & \myNocirc \\ 
\multicolumn{1}{l!{\vrule width 0.5pt}}{AvatarHunter~\cite{meng2023anonymization}} & \multicolumn{1}{c!{\vrule width 0.5pt}}{INFOCOM'23} & \multicolumn{1}{c!{\vrule width 0.5pt}}{Users' avatar} & \multicolumn{1}{c!{\vrule width 0.5pt}}{\cmark} & \multicolumn{1}{c!{\vrule width 0.5pt}}{\xmark} & \multicolumn{1}{c!{\vrule width 0.5pt}}{\xmark} & \multicolumn{1}{c!{\vrule width 0.5pt}}{\myNocirc} & \multicolumn{1}{c!{\vrule width 0.5pt}}{\myYescirc} & \myNocirc \\ 
\multicolumn{1}{l!{\vrule width 0.5pt}}{Yang~\etal~\cite{yang2024can}} & \multicolumn{1}{c!{\vrule width 0.5pt}}{Security'24} & \multicolumn{1}{c!{\vrule width 0.5pt}}{Users' avatar} & \multicolumn{1}{c!{\vrule width 0.5pt}}{\cmark} & \multicolumn{1}{c!{\vrule width 0.5pt}}{\xmark} & \multicolumn{1}{c!{\vrule width 0.5pt}}{\xmark} & \multicolumn{1}{c!{\vrule width 0.5pt}}{\myYescirc} & \multicolumn{1}{c!{\vrule width 0.5pt}}{\myNocirc} & \myNocirc \\
\multicolumn{1}{l!{\vrule width 0.5pt}}{GAZEploit~\cite{wang2024gazeploit}} & \multicolumn{1}{c!{\vrule width 0.5pt}}{CCS'24} & \multicolumn{1}{c!{\vrule width 0.5pt}}{User's avatar} & \multicolumn{1}{c!{\vrule width 0.5pt}}{\cmark} & \multicolumn{1}{c!{\vrule width 0.5pt}}{\xmark} & \multicolumn{1}{c!{\vrule width 0.5pt}}{\xmark} & \multicolumn{1}{c!{\vrule width 0.5pt}}{\myYescirc} & \multicolumn{1}{c!{\vrule width 0.5pt}}{\myYescirc} & \myNocirc \\
\multicolumn{1}{l!{\vrule width 0.5pt}}{Face-Mic~\cite{shi2021face}} & \multicolumn{1}{c!{\vrule width 0.5pt}}{MobiCom'21} & \multicolumn{1}{c!{\vrule width 0.5pt}}{Motion sensor} & \multicolumn{1}{c!{\vrule width 0.5pt}}{\cmark} & \multicolumn{1}{c!{\vrule width 0.5pt}}{\xmark} & \multicolumn{1}{c!{\vrule width 0.5pt}}{\cmark} & \multicolumn{1}{c!{\vrule width 0.5pt}}{\myYescirc} & \multicolumn{1}{c!{\vrule width 0.5pt}}{\myYescirc} & \myNocirc \\ 
\multicolumn{1}{l!{\vrule width 0.5pt}}{FaceReader~\cite{zhang2023facereader}} & \multicolumn{1}{c!{\vrule width 0.5pt}}{CCS'23} & \multicolumn{1}{c!{\vrule width 0.5pt}}{Motion sensor} & \multicolumn{1}{c!{\vrule width 0.5pt}}{\cmark} & \multicolumn{1}{c!{\vrule width 0.5pt}}{\xmark} & \multicolumn{1}{c!{\vrule width 0.5pt}}{\cmark} & \multicolumn{1}{c!{\vrule width 0.5pt}}{\myYescirc} & \multicolumn{1}{c!{\vrule width 0.5pt}}{\myYescirc} & \myNocirc \\
\multicolumn{1}{l!{\vrule width 0.5pt}}{BPSniff~\cite{ye2025bpsniff}} & \multicolumn{1}{c!{\vrule width 0.5pt}}{S\&P'25} & \multicolumn{1}{c!{\vrule width 0.5pt}}{Motion sensor} & \multicolumn{1}{c!{\vrule width 0.5pt}}{\cmark} & \multicolumn{1}{c!{\vrule width 0.5pt}}{\xmark} & \multicolumn{1}{c!{\vrule width 0.5pt}}{\cmark} & \multicolumn{1}{c!{\vrule width 0.5pt}}{\myYescirc} & \multicolumn{1}{c!{\vrule width 0.5pt}}{\myYescirc} & \myNocirc \\ \midrule
\multicolumn{2}{c!{\vrule width 0.5pt}}{\textbf{\sysname (Our work)}} & \multicolumn{1}{c!{\vrule width 0.5pt}}{Motion sensor} & \multicolumn{1}{c!{\vrule width 0.5pt}}{\cmark} & \multicolumn{1}{c!{\vrule width 0.5pt}}{\xmark} & \multicolumn{1}{c!{\vrule width 0.5pt}}{\cmark} & \multicolumn{1}{c!{\vrule width 0.5pt}}{\myYescirc} & \multicolumn{1}{c!{\vrule width 0.5pt}}{\myYescirc} & \myYescirc \\ \bottomrule
\end{tabular}%
\end{threeparttable}
\begin{tablenotes}
  \scriptsize
  \item $^{1}$~Three comparison metrics: \textbf{M1}--Malware installation; \textbf{M2}--Require external equipment (\eg, cameras, smartphones); and \textbf{M3}--Unrestricted sensor policy.
  \item $^{2}$``\myYescirc'' and ``\myNocirc'' indicate ``Yes'' and ``No'' that privacy leakage is explored.
\end{tablenotes}
\vspace{-0.2in}
\end{table}

\begin{figure*}[t]
    \minipage{0.48\textwidth}%
    \centering
      \begin{subfigure}[b]{.32\linewidth}
         \centering
         \includegraphics[width=\linewidth]{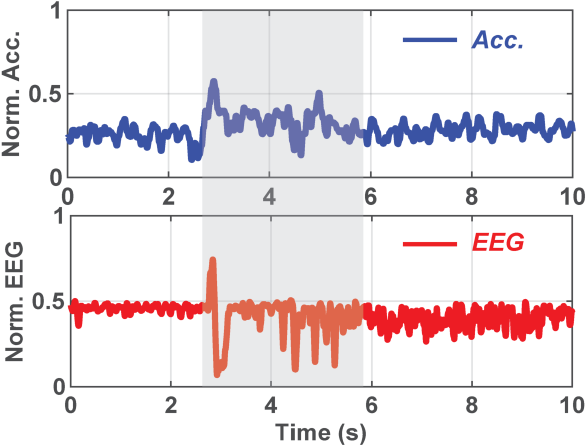}
         \vspace{-0.2in}
         \caption{Website.}
         \label{fig:signal_website}
    \end{subfigure}
    \begin{subfigure}[b]{.32\linewidth}
         \centering
         \includegraphics[width=\linewidth]{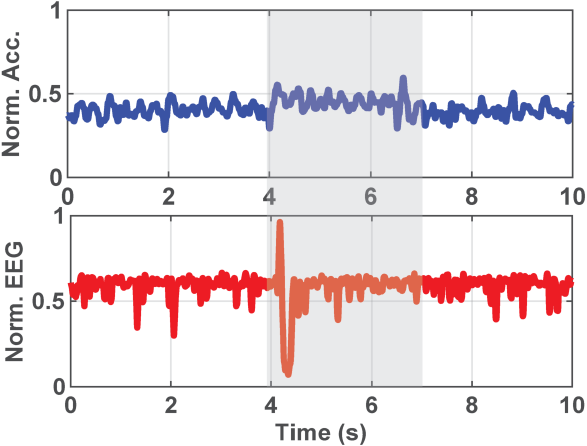}
         \vspace{-0.2in}
         \caption{App.}
         \label{fig:signal_app}
    \end{subfigure}
    \begin{subfigure}[b]{.32\linewidth}
         \centering
         \includegraphics[width=\linewidth]{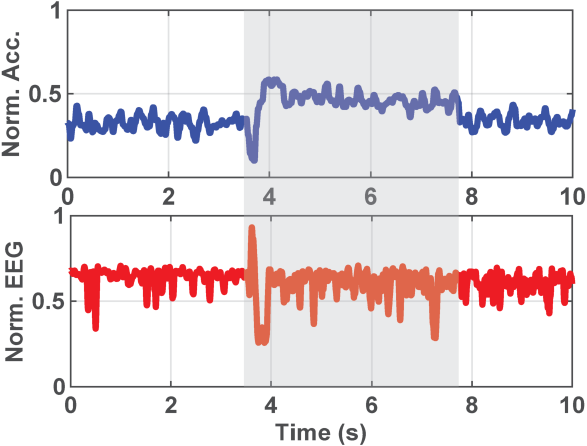}
         \vspace{-0.2in}
         \caption{Video.}
         \label{fig:signal_video}
    \end{subfigure}
      \vspace{-0.05in}
      \caption{UI-level visual presentations.}
      \label{fig:signal_ui_level}
    \endminipage\hfill
    \minipage{0.48\textwidth}%
    \centering
      \begin{subfigure}[b]{.32\linewidth}
         \centering
         \includegraphics[width=\linewidth]{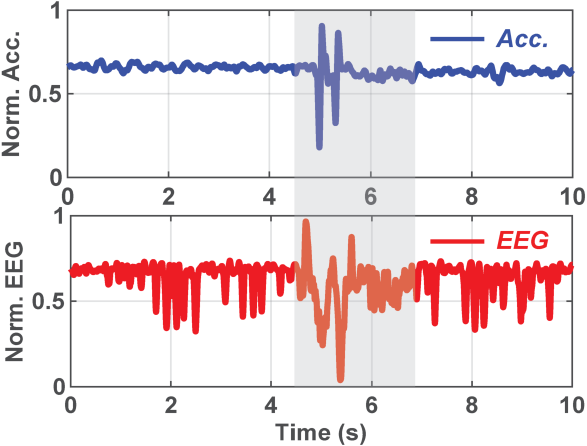}
         \vspace{-0.2in}
         \caption{Man \#1.}
         \label{fig:signal_male1}
    \end{subfigure}
    \begin{subfigure}[b]{.32\linewidth}
         \centering
         \includegraphics[width=\linewidth]{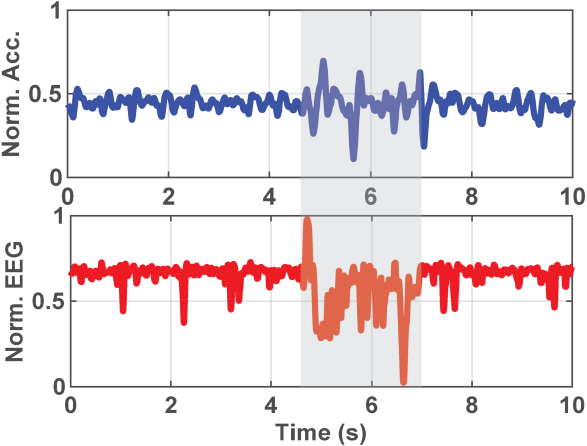}
         \vspace{-0.2in}
         \caption{Woman \#1.}
         \label{fig:signal_female1}
    \end{subfigure}
    \begin{subfigure}[b]{.32\linewidth}
         \centering
         \includegraphics[width=\linewidth]{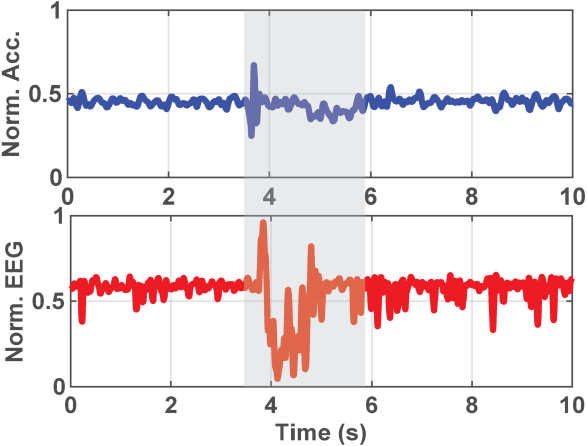}
         \vspace{-0.2in}
         \caption{Man \#2.}
         \label{fig:signal_male2}
    \end{subfigure}
      \vspace{-0.05in}
      \caption{User-level identity recognition.}
      \label{fig:signal_user_level_ui}
    \endminipage
    \vspace{0.05in}
    \minipage{0.64\textwidth}%
    \centering
      \begin{subfigure}[b]{.24\linewidth}
         \centering
         \includegraphics[width=\linewidth]{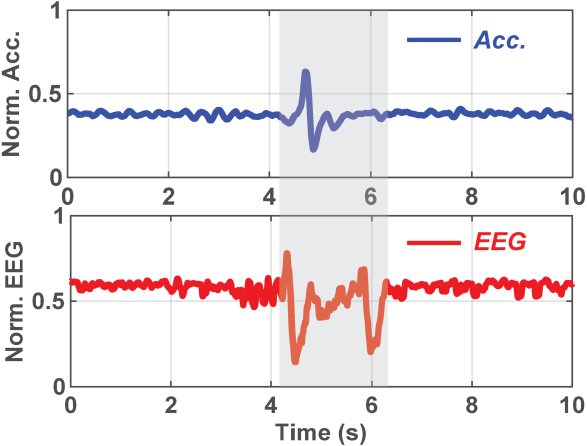}
         \vspace{-0.2in}
         \caption{Up.}
         \label{fig:signal_upper}
    \end{subfigure}
    \begin{subfigure}[b]{.24\linewidth}
         \centering
         \includegraphics[width=\linewidth]{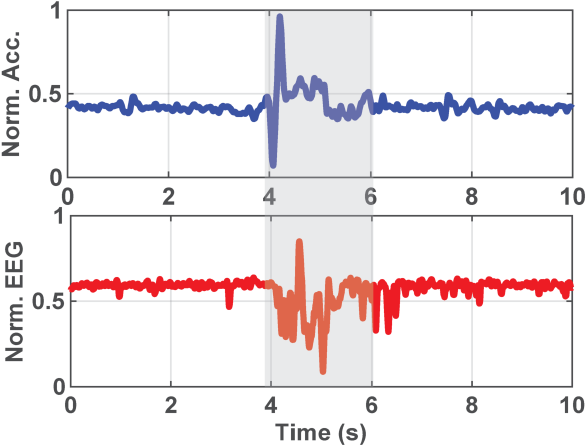}
         \vspace{-0.2in}
         \caption{Down.}
         \label{fig:signal_lower}
    \end{subfigure}
    \begin{subfigure}[b]{.24\linewidth}
         \centering
         \includegraphics[width=\linewidth]{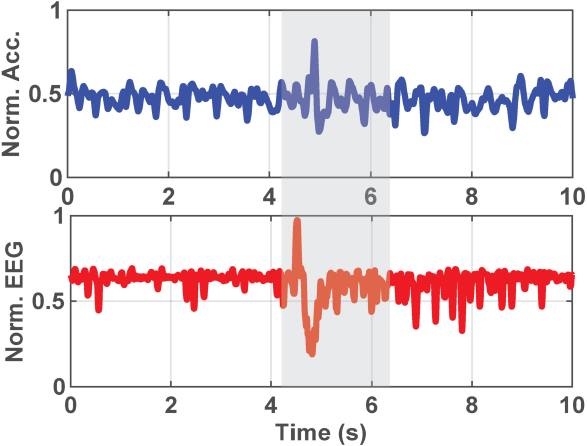}
         \vspace{-0.2in}
         \caption{Left.}
         \label{fig:signal_left}
    \end{subfigure}
    \begin{subfigure}[b]{.24\linewidth}
         \centering
         \includegraphics[width=\linewidth]{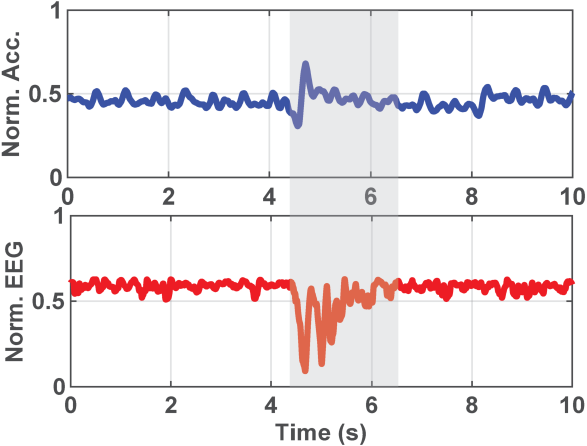}
         \vspace{-0.2in}
         \caption{Right.}
         \label{fig:signal_right}
    \end{subfigure}
      \vspace{-0.05in}
      \caption{UI-level reactions of gaze movements.}
      \label{fig:signal_user_level_ge}
    \endminipage\hfill
    \minipage{0.32\textwidth}%
    \centering
      \begin{subfigure}[b]{.48\linewidth}
         \centering
         \includegraphics[width=\linewidth]{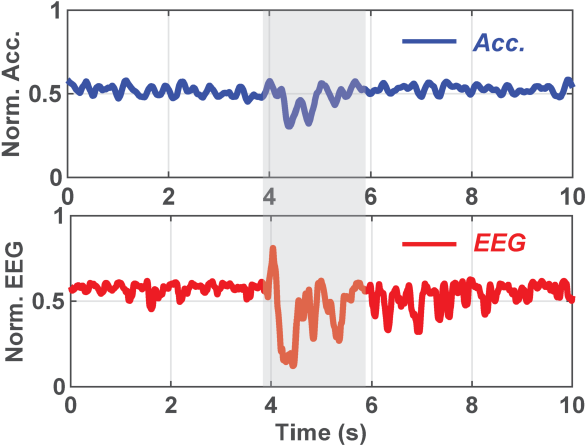}
         \vspace{-0.2in}
         \caption{Cat image.}
         \label{fig:signal_cat}
    \end{subfigure}
    \begin{subfigure}[b]{.48\linewidth}
         \centering
         \includegraphics[width=\linewidth]{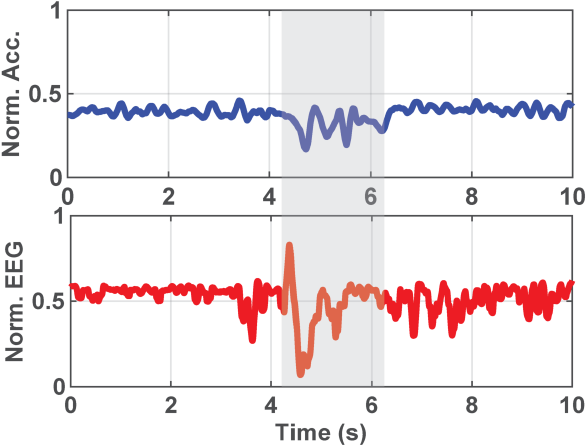}
         \vspace{-0.2in}
         \caption{Dog image.}
         \label{fig:signal_dog}
    \end{subfigure}
      \vspace{-0.05in}
      \caption{Brain-level image perception.}
      \label{fig:signal_brain_level}
    \endminipage
    \vspace{-0.15in}
\end{figure*}

As depicted in \autoref{fig:background_pupillary_response}, when users wear the VR headset, the integrated lenses display digital content within the virtual environment, triggering visual stimuli that lead to various pupillary responses.
\textit{(i) Static Pupillary Response:} When the eyes are stationary and focused on the displayed content (\eg, running apps, watching videos), the primary pupillary responses involve the dilation and contraction of the pupillary sphincter~\cite{slanzi2017combining, park2018infrared}.
In this scenario, pupillary responses to the content displayed in the VR headset are transmitted through the optic nerve to the brain for processing.
\textit{(ii) Dynamic Pupillary Response:} In contrast, when the users' eyes move during interactive activities in virtual scenes, such as searching for keys on a virtual keyboard or reading the text on the website, pupillary responses include both static changes and dynamic gaze information, which are detected by the ciliary ganglion and transmitted to the brain through the oculomotor nerve~\cite{zhang2023recognition}.

Both types of pupillary responses involve two distinct but correlated mechanisms. First, visual stimuli trigger pupillary sphincter dilation and contraction, producing subtle mechanical vibrations around the eyes~\cite{ye2025bpsniff} that propagate through the VR headset and are captured by the inbuilt motion sensors. Second, these pupillary responses independently correlate with brain EEG signals, as both are driven by the brain's visual processing pathway (\eg,~\cite{ko2020eyeblink, shahbakhti2021simultaneous, gusso2022more, slanzi2017combining, park2018infrared}). 
We note that the accelerometer measures mechanical vibrations rather than electrical neural activity directly, whereas the well-established statistical correlation between pupillary responses and EEG signals makes it feasible to reconstruct EEG-correlated representations from the captured vibrations. Based on this insight, as shown in \autoref{fig:leakage_model}, we model the correlations between the subtle vibrations in VR motion sensors and brain EEG variations, and systematically explore three levels of privacy inference targeting observable user behaviors and unobservable brain perceptions, including UI-level visual presentation, user-level identification and reaction, and brain-level image perceptions.
\looseness=-1

\section{Threat Model}
\label{sec:threat_model}

\subsection{Attack Vectors and Adversary Capability}
\label{subsec:adversary_capability}

Following the recent works~\cite{wu2023privacy, slocum2023going, luo2022holologger, zhang2023facereader, shi2021face, ye2025bpsniff, lee2025eyes} listed in \autoref{tab:comparison_vr_attacks},
we consider the potential attackers to be intentional VR application (app) developers who aim to acquire \textit{only} inbuilt unrestricted motion sensor data from the VR headset to 
infer sensitive user information
inside the virtual scenes.
Unlike other related works that rely on external equipment (\eg, Wi-Fi interceptor~\cite{al2021vr, khalili2026xr}, hidden cameras~\cite{gopal2023hidden, nguyen2024penetration}, charging cables~\cite{li2024dangers}, and microphones~\cite{luo2024eavesdropping}) or require privileged permission, such as access to system memory~\cite{zhang2023s} or virtual cameras~\cite{meng2023anonymization, yang2024can, wang2024gazeploit}, our assumption about the capability of the adversary is reasonable and practical.
This is because there are no deviations from the official VR app development guidelines specified by the associated SDKs (\eg, OpenXR SDK~\cite{picoopenxrsdk}, Oculus Mobile SDK~\cite{oculusmobilesdk}) that VR apps could access unrestricted motion sensor data without limits, including the inbuilt accelerometers and gyroscopes, from a user's headset in the background.
In the \textit{online deployment phase}, the adversary can only remotely obtain the motion sensor data from malicious code snippets embedded in these VR apps or third-party plugins released in open-source repositories.
As such, our hypothesized intentional app behaves in the same way as benign apps that are acceptable in official VR app stores.
Note that we do not assume that the adversary can access eye-tracking sensors to obtain gaze information directly because (1) most eye-tracking sensors are under higher restrictions and (2) only $17\%$ VR apps incorporate functionalities with eye movements~\cite{guo2024empirical}.
\looseness=-1

\subsection{Attack Surface and Objectives}
\label{subsec:attack_scenarios}

Unlike previous research, this work explores a new attack surface by measuring EEG signals derived from pupillary response-induced variations with unrestricted motion sensors,
despite previous studies using these motion sensors primarily for virtual keystroke inference~\cite{wu2023privacy, slocum2023going, shi2021face, zhang2023s, luo2022holologger}.
Furthermore, the attack objectives of this work not only involve unveiling a much finer granularity of a user's observable behaviors and privacy, which advances the current approaches' sole ability in coarse-grained privacy inference, but also extend beyond the scope of observable behaviors to unobservable behaviors.
Based on our three-level privacy model, the objectives are detailed as follows:
\looseness=-1

\paragraph{(L1) Observable UI-level Privacy}
Based on the static pupillary response, the adversary can use the reconstructed EEG signals from the victim's brain to infer 
sensitive information
related to various user-VR interactions, \ie, launching VR apps, browsing websites, and watching videos inside streaming apps like \textsc{Netflix}.
In particular, existing studies~\cite{wu2023privacy, slocum2023going, meteriz2022keylogging, luo2022holologger} cannot achieve UI-level privacy inference because they are based solely on motion sensor data, which lacks the necessary granularity to fingerprint specific apps, websites, and video content due to associated privacy sensitivities.
\autoref{fig:signal_ui_level} illustrates the correlation between motion sensor data (\eg, accelerometer, denoted as \textit{acc.}) and EEG signals during activities such as launching VR apps like \textsc{Netflix}, browsing websites such as \textsc{twitch.tv}, and watching videos like \textsc{3 Body Problem} on the Meta Oculus Quest 2 headset. The synchronous changes observed in both accelerometer and EEG signals confirm the feasibility of reconstructing brainwaves 
and inferring user-interface privacy.
\looseness=-1

\begin{figure*}[t]
    \centering
    \includegraphics[width=\linewidth]{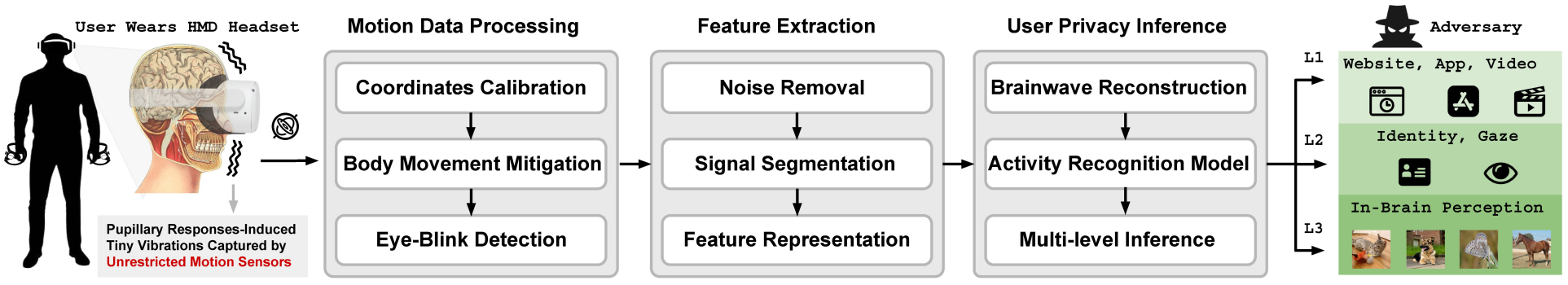}
    \vspace{-0.2in}
    \caption{Overview of \sysname.}
    \vspace{-0.2in}
    \label{fig:system_overview}
\end{figure*}

\paragraph{(L2) Observable User-level Privacy}
The adversary can exploit alterations in brain EEG signals to breach user-level privacy, such as identifying users to initiate de-anonymization attacks, a threat that has been widely recognized threats by prior studies~\cite{meng2023anonymization, shi2021face, zhang2023facereader}. For example, \autoref{fig:signal_user_level_ui} displays reconstructed EEG signals from three different VR users (two men and one woman).
Moreover, based on the dynamic pupillary response (\autoref{subsec:background_pupillary}), adversaries can analyze reconstructed brainwaves to track the gaze movements of VR users. This capability allows them to connect user attention data to more accurately determine UI-level privacy concerns, such as virtual keystrokes, by monitoring user-level gaze movements toward specific keys on a virtual keyboard. \autoref{fig:signal_user_level_ge} illustrates the alterations in accelerometer and EEG signals with gaze movements: up, down, left, and right, respectively.
As a result, these examples demonstrate the feasibility of using reconstructed EEG signals for user-level identity recognition and gaze movement extraction.

\paragraph{(L3) Unobservable Brain-level Privacy}
The adversary leverages data collected from unrestricted motion sensor to reconstruct EEG signals for inferring brain-level visual perceptions, such as decoding perceptive images within the user's brain, which is a well-established brain-level task and has been widely validated by previous EEG-to-image studies (\eg, ~\cite{bai2023dreamdiffusion, lan2023seeing, kavasidis2017brain2image, tirupattur2018thoughtviz, davis2022brain}).
Specifically, the adversary uses the reconstructed EEG signals to fine-tune advanced generative models, \ie, DreamDiffusion\cite{bai2023dreamdiffusion}, which interpret the EEG data to match the images perceived in the user's brain.
\autoref{fig:signal_brain_level} exhibits the accelerometer and EEG signals when the VR user visualizes images in the mind, demonstrating that distinct perceptive images generate unique EEG patterns, which highlight the potential to uncover brain-level visual perceptions within the current threat model.


\section{Attack Design}
\label{sec:attack_design}

\subsection{Overview of \sysname}
\label{subsec:system_overview}

\autoref{fig:system_overview} shows the end-to-end system overview of \sysname.
The adversary first obtains the unrestricted motion sensor data from the victim's VR headset and then conducts the following three stages: \textit{(1) Motion Data Processing:} The adversary leverages the acquired motion data to calibrate the initial coordinates, mitigate the influence resulting from body movement, and detect the eye-blink events;
\textit{(2) Feature Extraction:} Then, the adversary applies filters to remove high-frequency noise, divide the informative segments from the signals, and extract feature representations;
and \textit{(3) User Privacy Inference:} The extracted features are then used to reconstruct brain wave signals (EEG signals), which are then used to infer various user privacy at different granularity levels.
Finally, the adversary could obtain the output to 
infer observable sensitive information
at the UI level (\eg, website history, app usage, video preference), user level (\eg, identity, gaze movements, and keystrokes), and unobservable brain level (\eg, in-brain perceptive images).
\looseness=-1

\subsection{Motion Data Processing}
\label{subsec:motion_data_processing}

\paragraph{Coordinates Calibration} To mitigate the impact of varying initial body postures, we first perform coordinate calibration on the collected motion sensor data.
Since interacting with a VR headset involves inherently three-dimensional movements, the 3D motion data captured by inbuilt accelerometers, which are recorded independently at different body postures, lack spatial coordinate calibration, making it unsuitable for direct use in brainwave reconstruction.
To resolve this, we transform the acceleration values associated with different head postures into a common body reference coordinate system that is independent of orientation and location. We define the world coordinate system by the axes north, east, and down, or in the direction of gravity, and refer to the local coordinate system of the VR headset as $(X, Y, Z)$. The plane of motion for the headset is defined as the front-side plane, which is perpendicular to gravity, with the side pointing toward the right side of the user's forehead
Assuming the linear acceleration signals along the three orthogonal directions of the VR headset are $Acc_{X}$, $Acc_{Y}$, and $Acc_{Z}$,  we compute the linear acceleration in the reference system as follows:
\begin{equation}
\small
\begin{bmatrix}
Acc_{X^{'}}\\ 
Acc_{Y^{'}}\\ 
Acc_{Z^{'}}
\end{bmatrix}=R_{b}^{w}\cdot R_{w}^{v}\cdot \begin{bmatrix}
Acc_{X}\\ 
Acc_{Y}\\ 
Acc_{Z}
\end{bmatrix}
\end{equation}

Specifically, $Acc_{X^{'}}$, $Acc_{Y^{'}}$, and $Acc_{Z^{'}}$ are linear accelerations along the direction of gravity, the front direction, and the side direction.
The transformation matrices $R_{h}^{w}$ and $R_{w}^{v}$, which represent the rotation from the world coordinate system to the body coordinate system and from the VR headset coordinate system to the world coordinate system, can be derived using angular data from the inbuilt motion sensors, based on established methods in previous work~\cite{mohssen2014s}.
Note that accurately determining the absolute head direction of a VR user is not feasible using only the inbuilt accelerometer of a VR headset. Therefore, our focus is primarily on coordinate calibration rather than precise real-world head posture.
Once the acceleration data are transformed into the body coordinate system, we use $Acc_{X^{'}}$, $Acc_{Y^{'}}$, and $Acc_{Z^{'}}$ as calibrated motion sensor data to proceed with further analysis.
In addition, as shown above, the $Acc_{Y^{'}}$ presents similar signal patterns to EEG signals in response to user activities at different levels, which we select to reconstruct brain EEG signals for further analysis.
\looseness=-1

\begin{figure}[t]
    \begin{subfigure}[b]{.495\linewidth}
         \centering
         \includegraphics[width=\linewidth]{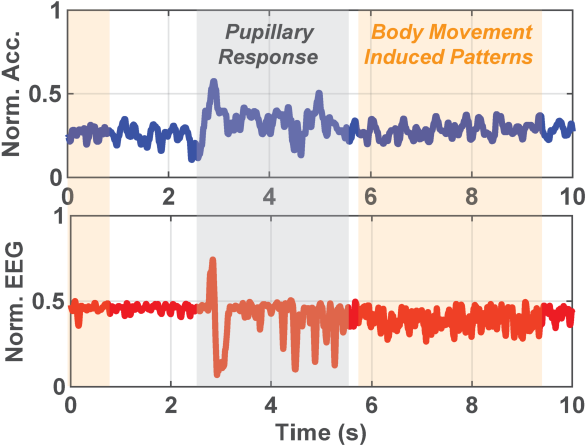}
         \vspace{-0.2in}
         \caption{Raw signals.}
         \label{fig:before_bmm}
    \end{subfigure}
    \begin{subfigure}[b]{.495\linewidth}
         \centering
         \includegraphics[width=\linewidth]{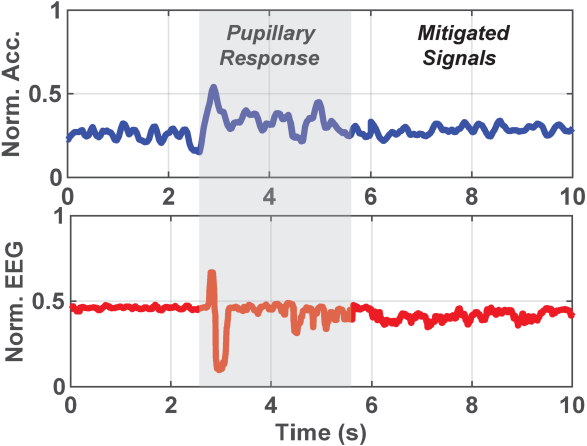}
         \vspace{-0.2in}
         \caption{Body movement mitigation.}
         \label{fig:after_bmm}
    \end{subfigure}
    \vspace{-0.2in}
    \caption{Recorded acceleration and EEG signals before and after body movement mitigation.}
    \vspace{-0.2in}
    \label{fig:bmm}
\end{figure}

\paragraph{Body Movement Mitigation} 
In practical settings, users generally interact with the VR headset with inevitably subtle body movements.
Hence, the embedded motion sensors in the VR headset could capture diverse and unpredictable motion-related patterns, \ie, spontaneous head movements and instinctive body rotations, which introduce extra interference in the collected motion sensor data due to the overlap in the frequency range associated with accelerations.
To effectively reduce these noise patterns, we have implemented a generalized \textit{Short-Time Energy} (STE) approach to realize adaptive time-series filtering~\cite{zhang2023facereader, li2023adaptive}.
We apply a two-second sliding window to the obtained acceleration signal $\mathcal{S}(t)$ and then calculate the total energy of this segment, which is used to determine the presence of significant body motions in the analyzed signals.
Specifically, the obtained segment energy is compared with a power threshold, and an adaptive filter is applied when the detected segment energy exceeds the empirically defined power threshold $\mathcal{T}$ ($0.5$).
To obtain the suitable configuration of the adaptive filter and to acquire the filtered sensor signals, assuming the reference signal $\mathcal{S}_{R}(t)$ is collected when the user's head is absolutely static, we have to solve this optimization problem by establishing the optimal adaptive weight vector $\mathcal{V}_{w}$ as follows:
\looseness=-1
\begin{equation}
\small
\begin{aligned}
& \underset{\mathcal{V}_{w}}{\text{argmin}}
& & \sum_{t \in \tau} D_{KL}(\mathcal{S}_{R}(t), \mathcal{S}(t)) = \sum_{t \in \tau} \mathcal{S}_{R}(t) \log \left( \frac{\mathcal{S}_{R}(t)}{\mathcal{S}(t)} \right) \\
& \text{subject to}
& & \sum_{t \in \tau} \mathcal{V}_{W}(t) = \alpha \sum_{t \in \tau} \mathcal{V}_{W}(t) + \delta \sum_{t \in \tau} \mathcal{E}(t) \cdot \mathcal{S}^{'}(t), \\
& & & \sum_{t \in \tau} \mathcal{S}(t) = \sum_{t \in \tau} \mathcal{V}_{W}(t) \cdot \mathcal{S}^{'}(t), \sum_{t \in \tau} \mathcal{S}^{'}(t)^2 \leq \mathcal{T}.
\end{aligned}
\end{equation}

In particular, $\mathcal{S}^{'}(t)$ and $\mathcal{E}(t)$ represent the filtered signals after body movement mitigation and the error function, and $\tau$, $\alpha$, and $\delta$ are the indices of the time-series sensor signal, hyper-parameters, and the optimizing step size, respectively.
In addition, we utilize the \textit{Kullback-Leibler} (KL) divergence~\cite{huang2019novel} as the error function to measure discrepancies between the unfiltered and reference signals and to optimize the weights of the adaptive filter because the KL divergence presents a robust ability to quantify differences between the distributions of two series of signals.
In particular, \autoref{fig:before_bmm} and \autoref{fig:after_bmm} illustrate the acceleration and EEG signals before and after applying the body movement mitigation. We can observe that the patterns of interference resulting from subtle body movements are significantly mitigated.
\looseness=-1

\begin{figure}[t]
    \begin{subfigure}[b]{.495\linewidth}
         \centering
         \includegraphics[width=\linewidth]{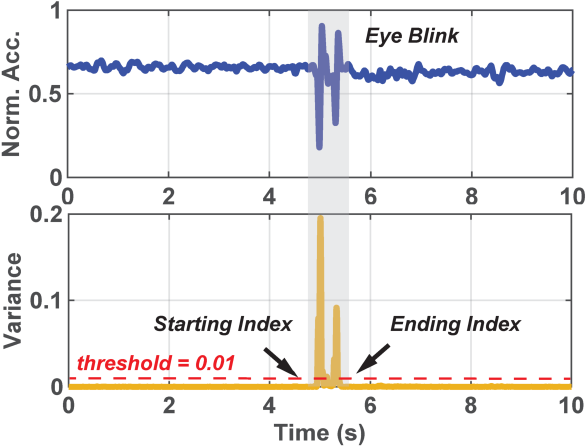}
         \vspace{-0.2in}
         \caption{Eye-blink segmentation.}
         \label{fig:eb_acc_var}
    \end{subfigure}
    \begin{subfigure}[b]{.495\linewidth}
         \centering
         \includegraphics[width=\linewidth]{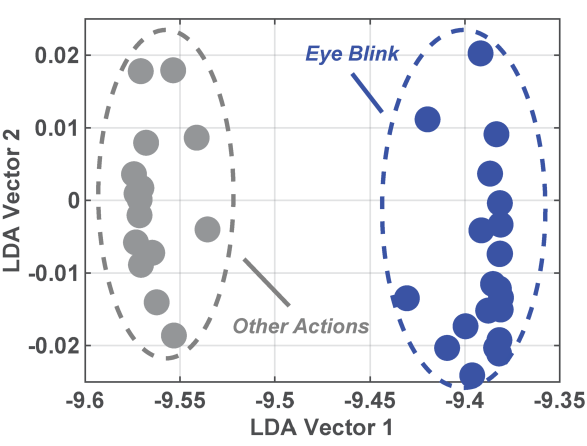}
         \vspace{-0.2in}
         \caption{LDA distribution.}
         \label{fig:eb_distribution}
    \end{subfigure}
    \vspace{-0.15in}
    \caption{Eye-blink detection, including variance-based signal segmentation and LDA distributions}
    \vspace{-0.15in}
    \label{fig:eb_acc_distribution}
\end{figure}

\paragraph{Eye-Blink Detection} During user interactions with a VR headset, involuntary eye blinks commonly occur.
Previous studies (\eg, \cite{ko2020eyeblink, shahbakhti2021simultaneous}) have shown that these eye blinks could induce dynamic pupillary responses, leading to detectable vibrations in motion sensor data and EEG signals.
These EEG responses to eye blinks vary between individual users because of differences in biometric structure, depicting a potential method for user de-anonymization. To address this, we developed an eye-blink detector that identifies eye blinks from the acceleration data captured by the VR headset's unrestricted motion sensors.
\autoref{fig:eb_acc_var} illustrates the significant changes in acceleration signals triggered by eye blinks while wearing a Meta Oculus Quest 2 headset. Typically, one eye blink lasts between \SI{0.2}{\second} and \SI{2}{\second}~\cite{shahbakhti2021simultaneous}.
Hence, in practice, we determine the starting and ending indices of an eye-blink event by applying the moving variance window with a length of \SI{0.2}{\second} and an empirical threshold of $0.01$.
Furthermore, we calculate the \textit{Linear Discriminant Analysis} (LDA)~\cite{xanthopoulos2013linear} values of the amplitude within segments of eye-blink and other activities, and use a k-NN detector to accurately identify eye blinks, as detailed in \autoref{fig:eb_distribution}, which achieves $100\%$ accuracy in determining eye blinks from the obtained acceleration signals.
\looseness=-1

\subsection{Feature Extraction}
\label{subsec:feature_extraction}

\paragraph{Noise Removal} The accelerometers embedded in headsets are susceptible to high-frequency interference from human speaking frequencies (\eg, $90$--\SI{255}{\hertz}) and high-frequency electromagnetic radiation at kHz and MHz levels.
Therefore, to mitigate the influence of this interference, we apply a \textit{Savitzky-Golay} (S-G)~\cite{chen2004simple} filter to the collected time-series signals, which could remove this high-frequency noise without distorting the signal shapes.
In practice, we set the window size to $0.1$ of the sampling rate and the polynomial order to $3$.
Then, considering the different starting amplitude values of acceleration signals, we calculate the average value of the first one-second data as the static acceleration data and then deduct this offset from the motion sensor data to correct for any biases.
\looseness=-1

\paragraph{Signal Segmentation} After obtaining the filtered signals, we then divide the signal segments that contain relevant user activities and information.
As discussed in \autoref{subsec:motion_data_processing}, burst fluctuations in acceleration signals due to pupillary responses can be identified through variance analysis. To this end, we utilize a moving variance window with a threshold of $0.05$ applied to the filtered signals.
This approach allows us to detect pupillary responses at the appearance of the first peak. Subsequently, the segment of acceleration data between the first and second peaks in the moving variance signal will be extracted because it contains critical user activity information.
In addition, due to the variability in the duration of different pupillary responses (\eg, eye blink: $0.2$--\SI{2}{\second}, opening a VR app: $1$--\SI{5}{\second}), we normalize the processed signals of each attempt to the same length of time (\eg, \SI{0.5}{\second}) by exploiting up-sampling (\eg, interpolation~\cite{rukundo2012nearest}) or down-sampling (\eg, decimation factor~\cite{kreindler2016effects}) algorithms.
\looseness=-1

\begin{figure}[t]
    \begin{subfigure}[b]{.495\linewidth}
         \centering
         \includegraphics[width=\linewidth]{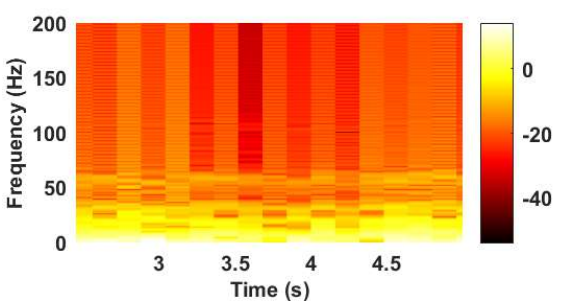}
         \vspace{-0.2in}
         \caption{Accelerometer.}
         \label{fig:spectrogram_acc}
    \end{subfigure}
    \begin{subfigure}[b]{.495\linewidth}
         \centering
         \includegraphics[width=\linewidth]{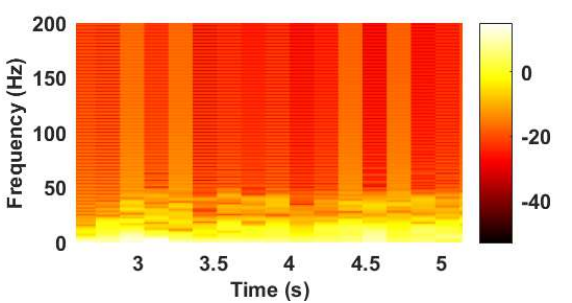}
         \vspace{-0.2in}
         \caption{Brain EEG.}
         \label{fig:spectrogram_eeg}
    \end{subfigure}
    \vspace{-0.15in}
    \caption{Spectrograms of acceleration and EEG signals.}
    \vspace{-0.15in}
    \label{fig:feature_representations}
\end{figure}

\paragraph{Feature Representation} 
To explore the interconnected features between the acceleration and EEG signal segments comprehensively, we apply the \textit{Short-Term Fourier Transform} (STFT) to these time-series signals, extracting time-frequency spectrogram images as feature representations. We assume that the sampling frequency for the inbuilt accelerometer is $f_{acc}$ and that the processed signal segments are $\overline{\mathcal{S}^{'}}(t)$.
The FFT of the acceleration signals is given by:

\begin{equation}
\small
    \overline{\mathcal{S}^{'}}(f) = STFT(\overline{S^{'}}(t)) = \int_{0}^{t_{s}} S(t)\cdot\omega (t-\sigma)\cdot e^{-i2\pi ft} dt
\end{equation}
Specifically, we use a hamming window of size $0.5f_{acc}$ with an overlapping rate of $50\%$ to convert the segment into discrete expressions.
The terms $\omega$ and $\sigma$ represent the window function and the time around the center of the window, respectively.
For ground truth EEG signals captured at a different sampling frequency $f_{eeg}$, we adjust the sliding window size to $0.5f_{eeg}/f_{acc}$ and convert the time-series EEG signals to a spectrogram with the same time frame length.
\autoref{fig:spectrogram_acc} and \autoref{fig:spectrogram_eeg} illustrate the spectrograms for the acceleration and EEG signals, respectively. These spectrogram images, particularly those related to pupillary response segments shown in \autoref{fig:after_bmm}, exhibit similar pattern distributions, validating the potential of using acceleration signals to reconstruct brain EEG signals. Hence, these feature representations from both signal types are utilized as inputs and outputs for the brainwave reconstruction model, which will be illustrated in subsequent sections.

\subsection{User Privacy Inference}
\label{subsec:user_privacy_reconstruction}

\paragraph{Brainwave Reconstruction} To reconstruct the brainwave, we first build a \textit{Conditional Generative Adversarial Network} (cGAN) model to generate the brain EEG spectrogram $\mathcal{P}_{E}$ using the given acceleration spectrogram $\mathcal{A}$ as conditions.
The cGAN model extracts the mapping relationship from the acceleration spectrogram to the corresponding EEG spectrogram and converts it to time-series brain EEG signals to further uncover VR user privacy.
In practice, we leverage the pix2pix~\cite{isola2017image} image-to-image generative model as the backbone, which has shown remarkable capabilities in transformations between paired training data.
Specifically, the cGan model consists of two primary components: a generator $\mathcal{G}_{\theta}$ and a discriminator $\mathcal{D}_{\phi}$.
The generator is tasked with generating brain EEG spectrogram $\mathcal{P}_{E}$ that is indistinguishable from the acceleration spectrogram $\mathcal{P}_{A}$ given the inbuilt accelerometer data from the VR headset.
On the other hand, the discriminator tries to distinguish between $\mathcal{P}_{E}$ and $\mathcal{P}_{A}$.
The generator $\mathcal{G}_{\theta}$ takes the acceleration spectrogram $\mathcal{A}$ as input and generates a reconstructed spectrogram $\mathcal{P}_{E}=\mathcal{G}_{\theta}(\mathcal{A})$.
The discriminator $\mathcal{D}_{\phi}$ receives the groundtruth image pair $(\mathcal{P}_{A}, \mathcal{A})$ and a reconstructed image pair $(\mathcal{P}_{E}, \mathcal{A})$ and attempts to fit the EEG signals from the brainwave sensors.
The training process involves alternating between updating the parameters and weights of the generator $\mathcal{G}_{\theta}$ and the discriminator $\mathcal{D}_{\phi}$.
In particular, $\mathcal{D}_{\phi}$ is trained to maximize its ability to correctly fit the real and generated EEG spectrograms, while $\mathcal{G}_{\theta}$ is trained against $\mathcal{D}_{\phi}$ to obtain optimal efficacy.

To realize the training of this cGAN model, we design and implement the objective loss function $\mathcal{L}(\theta, \phi)$ considering both adversarial loss and L1 loss.
In particular, adversarial loss ensures that the generated EEG spectrograms are close to those of the EEG signals from brainwave sensors, while L1 loss ensures fidelity to the input acceleration profiles. Hence, the objective loss function is shown as:
\begin{equation}
\small 
    \mathcal{L}(\theta, \phi)=\mathcal{L}_{adv}(\mathcal{G}_{\theta}, \mathcal{D}_{\phi}) + \lambda\mathcal{L}_{L_1}(\mathcal{G}_{\theta})
\end{equation}
where $\mathcal{L}_{adv}$, $\mathcal{L}_{L1}$, and $\lambda$ are the adversarial loss, L1 loss, and the hyperparameter for balancing the two losses, respectively.
Specifically, the adversarial loss $\mathcal{L}_{adv}$ and L1 loss $\mathcal{L}_{L1}$ can be formulated by considering the acceleration and EEG spectrograms as follows:

\begin{equation}
\small
\begin{cases}
\begin{split}
    \mathcal{L}_{adv}(\mathcal{G}_{\theta}, \mathcal{D}_{\phi}) &= \mathbb{E}_{\mathcal{P}_{A}, \mathcal{A}\sim p(\mathcal{P}_{A}, \mathcal{A})}\log \mathcal{D}_{\phi} (\mathcal{P}_{A}, \mathcal{A}) \\
    & + \mathbb{E}_{\mathcal{A}\sim p(\mathcal{A}), \mathcal{P}_{E}\sim \mathcal{G}_{\theta}(\mathcal{A})}\log (1-\mathcal{D}_{\phi} (\mathcal{P}_{E}, \mathcal{A}))
\end{split} \\
\mathcal{L}_{L1}(\mathcal{G}_{\theta}) = \mathbb{E}_{\mathcal{P}_{A}, \mathcal{A}\sim p(\mathcal{P}_{A}, \mathcal{A}), \mathcal{P}_{E}\sim \mathcal{G}_{\theta}(\mathcal{A})}\left | \mathcal{P}_{A}-\mathcal{P}_{E} \right |
\end{cases}
\end{equation}
where we set the batch size to one and used the Adam optimizer with $0.5$ momentum rate in both the generator $\mathcal{G}_{\theta}$ and the discriminator $\mathcal{D}_{\phi}$, with an initial learning rate of 0.0001 and an initial balance parameter $\lambda = 100$.
In addition, we utilize the U-Net~\cite{ronneberger2015u} as the backbone in $\mathcal{G}_{\theta}$ and resample both the acceleration and EEG spectrograms to a size of $256\times 256$ and trained the cGAN model for $1000$ epochs.
\looseness=-1

\begin{figure}[t]
    \begin{subfigure}[b]{.495\linewidth}
         \centering
         \includegraphics[width=\linewidth]{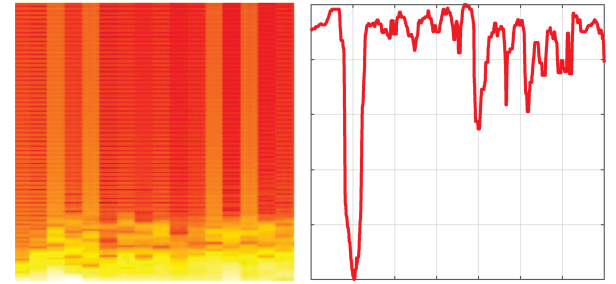}
         \vspace{-0.15in}
         \caption{Original EEG.}
         \label{fig:original_spec_signal}
    \end{subfigure}
    \begin{subfigure}[b]{.495\linewidth}
         \centering
         \includegraphics[width=\linewidth]{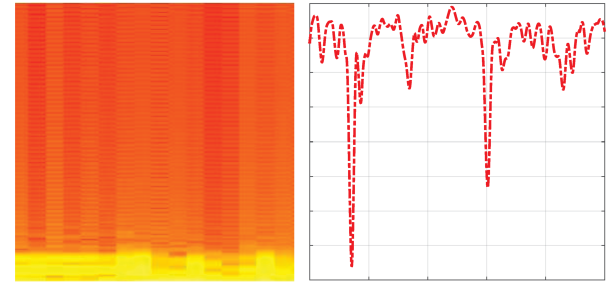}
         \vspace{-0.15in}
         \caption{Reconstructed EEG.}
         \label{fig:reconstructed_spec_signal}
    \end{subfigure}
    \vspace{-0.15in}
    \caption{Original and reconstructed time-frequency EEG spectrograms and time-series EEG signals.}
    \vspace{-0.2in}
    \label{fig:original_reconstructed_eeg}
\end{figure}

\autoref{fig:reconstructed_spec_signal} shows an example of reconstructed EEG spectrograms using acceleration spectrograms. Specifically, the reconstructed EEG spectrograms are leveraged to train different activity recognition models to infer UI-level, user-level activity, and privacy.
In addition, we apply the \textit{Inverse Short-Term Fourier Transform} (ISTFT) to reconstruct the EEG signals to fit the real brainwave signal as:
\looseness=-1

\begin{equation}
\small
    \overline{S^{''}}(t) = \frac{1}{\int_{0 }^{t_{s}}\left | \omega (t) \right |^{2}dt}\int_{0 }^{t_{s}}\int_{0 }^{t_{s}}\overline{\mathcal{S}^{'}}(f)\cdot \omega (t-\sigma)\cdot e^{i2\pi ft}dfd\sigma
\end{equation}

In addition, \autoref{fig:original_reconstructed_eeg} also presents an example of original and reconstructed EEG signals. We can observe that while most informative patterns are preserved, some information loss occurs due to the signal processing and compression steps involved in STFT and ISTFT. Despite this, the reconstructed EEG signal remains effective as input to fine-tune EEG-to-image models, \ie, DreamDiffusion~\cite{bai2023dreamdiffusion}, to infer the brain-level visual perceptions of VR users.
\looseness=-1

\paragraph{Activity Recognition Model} In particular, we design and implement different activity recognition models to infer the aforementioned three-level user privacy when using the VR headset.
First, to infer observable UI-level (\eg, website history, app usage) and user-level (\eg, identification, gaze movements) privacy, we design a transformer-based activity recognition model that takes the reconstructed EEG spectrogram images as input and recognizes the corresponding user activities.
Specifically, the model architecture is designed based on the \textit{Vision Transformer} (ViT)~\cite{dosovitskiy2020image}, which consists of an input layer that resizes and normalizes the input image, a patch embedding layer to split the image into $16\times 16$ patches, a positional encoding to add fixed or learnable positional embeddings to each patch for positional information retention, a transformer encoder with $12$ layers of multi-head self-attention and feed-forward networks, and a classification layer that calculates the average pooling over the output and uses a fully-connected layer to map the transformer output to the class with the highest probability.
We implement the activity recognition model in PyTorch 2.0 with an initial learning rate of $0.001$ and cross-entropy loss as the loss function. The output shape of the last fully-connected layer depends on the corresponding task (\eg, the number of websites, apps, and videos).
\looseness=-1


Second, to rebuild unobservable brain-level image perceptions, we leverage the open-sourced MOABB dataset~\cite{jayaram2018moabb}, which represents the largest available open-sourced EEG-image dataset, to train an EEG encoder that generates valid input for fine-tuning the pre-trained Stable Diffusion model proposed in DreamDiffusion~\cite{bai2023dreamdiffusion}. 
Specifically, we replace data in $40$ common classes with the EEG signals and images collected by our experimental devices (\autoref{subsec:experimental_setup}) and train the EEG encoder with a mask ratio of $0.75$ over $800$ epochs, which we validate as a classification task rather than pixel-faithful image reconstruction.
Next, the cross-attention heads and the trained EEG encoder will be jointly optimized with EEG-image pairs to accomplish fine-tuning.
The loss function in the Stable Diffusion fine-tuning process can be described as follows:

\begin{equation}
\small
    \mathcal{L}_{SD} = \mathbb{E}_{x, \epsilon\sim \mathcal{N}(0, 1), t}\left \| \epsilon-\epsilon^{'}(x_{t}, t, \tau^{'}(y) \right \|^{2}_{2}
\end{equation}
where $x$ is the given image, $y$ is the output of the EEG encoder, $\tau^{'}$ is the projector to obtain the embedding EEG representations, and $\epsilon^{'}$ is the denoising function implemented as the U-Net.

\begin{figure}[t]
    \begin{subfigure}[b]{.495\linewidth}
         \centering
         \includegraphics[width=\linewidth]{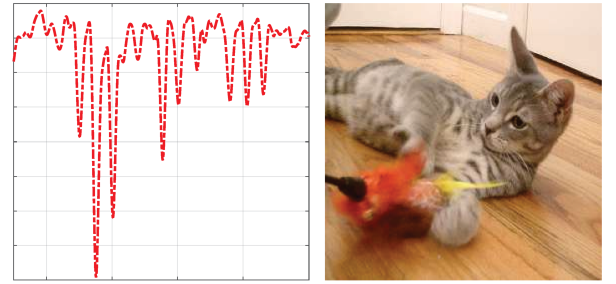}
         \vspace{-0.15in}
         \caption{Reconstructed EEG of Cat.}
         \label{fig:reconstructed_cat}
    \end{subfigure}
    \begin{subfigure}[b]{.495\linewidth}
         \centering
         \includegraphics[width=\linewidth]{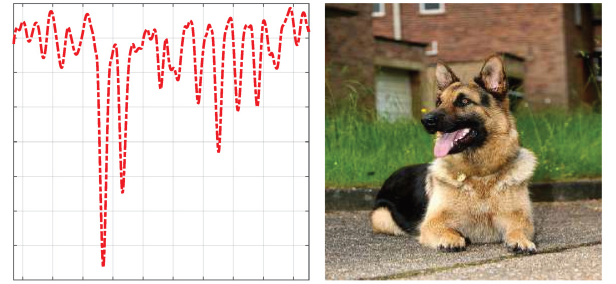}
         \vspace{-0.15in}
         \caption{Reconstructed EEG of Dog.}
         \label{fig:reconstructed_dog}
    \end{subfigure}
    \vspace{-0.15in}
    \caption{Reconstructed EEG signals and brain-level visual perception of a ``Cat'' image and a ``Dog'' image.}
    \vspace{-0.2in}
    \label{fig:reconstructed_images}
\end{figure}

In the end, the fine-tuned Stable Diffusion model will take the reconstructed EEG signal as input and generate the image prediction that represents the visual perceptions inside the user's brain.
For instance, \autoref{fig:reconstructed_cat} and \autoref{fig:reconstructed_dog} present the reconstructed EEG signals and the corresponding brain-level image perception of ``Cat'' and ``Dog'', respectively.
Note that due to the versatile and complex characteristics of human eye perception and brain thought, we take perceptive images as an example to demonstrate the EEG-to-image perception capability of \sysname in this paper, whose feasibility has been widely validated and recognized by most previous relevant studies (\eg, \cite{bai2023dreamdiffusion, lan2023seeing, kavasidis2017brain2image, tirupattur2018thoughtviz, davis2022brain}).
\looseness=-1

\paragraph{Multi-level Inference} Ultimately, the adversary can exploit \sysname to identify three levels of user activities while wearing a VR headset, thus exposing user privacy and sensitive information. At the UI level, well-documented privacy violations, such as fingerprinting on the website and app~\cite{panchenko2016website} and detailed streaming video leakage~\cite{schuster2017beauty} can lead to the disclosure of metadata related to political and financial affiliations, potentially resulting in credential leakage.
Furthermore, \sysname facilitates UI-level activity recognition, which previous efforts could not achieve using only unrestricted motion sensors in the VR headset (\eg, \cite{luo2022holologger, slocum2023going, wu2023privacy}).
At the user level, \sysname can utilize the reconstructed EEG signals to de-anonymize the VR user's identity and recognize gaze movements that might cause credential breaches, such as keystroke inference~\cite{wang2019your, chen2018eyetell, abdrabou2022your, ni2023uncovering}.
Finally, \sysname is the first work that reveals brainwave-aware user privacy, \ie, unveiling visual perceptive images formed in the brain from reconstructed EEG signals.



\section{Evaluation}
\label{sec:evaluation}

\looseness=-1

\subsection{Experimental Setup}
\label{subsec:experimental_setup}

\paragraph{Experiment Devices} \sysname is evaluated on four commercial VR devices, \ie, Meta Oculus Quest 2, Meta Oculus Quest, PICO 4 All-in-One, and HTC Vive Pro. In particular, the Meta Oculus Quest 2 and Meta Oculus Quest are equipped with a motion sensor board (\ie, 330-00193-03 1PASF8K~\cite{zhang2023facereader}) originally designed by Meta. The PICO 4 All-in-One adopts the TDK ICM42688 motion tracking module~\cite{motionsensorpico}, and the HTC Vive Pro contains a G-sensor~\cite{htcvivespec}, which consists of an accelerometer and a gyroscope.
In practice, we set the sampling frequency of the motion sensors in the four VR headsets as \SI{500}{\hertz}, which is the most stable sampling frequency for these headsets.
We developed a tool to collect motion sensor data in the background from Meta Oculus Quest 2 and Meta Oculus Quest through the function \textit{ovr\_GetTrackingState()} on Oculus Mobile SDK~\cite{oculusmobilesdk}, as well as implemented a tool for data collection in PICO 4 All-in-One and HTC Vive Pro through the function \textit{getDeviceToAbsoluteTrackingPose()} in OpenVR SDK~\cite{openvr} or \textit{getViewerPose()} in the WebXR Device API~\cite{webxrapi}.
\autoref{fig:code_snippets} in the Appendix depicts the code snippets implemented in the data collection tools on different VR platforms.
\looseness=-1

\begin{figure}[t]
    \begin{subfigure}[b]{.495\linewidth}
         \centering
         \includegraphics[width=\linewidth]{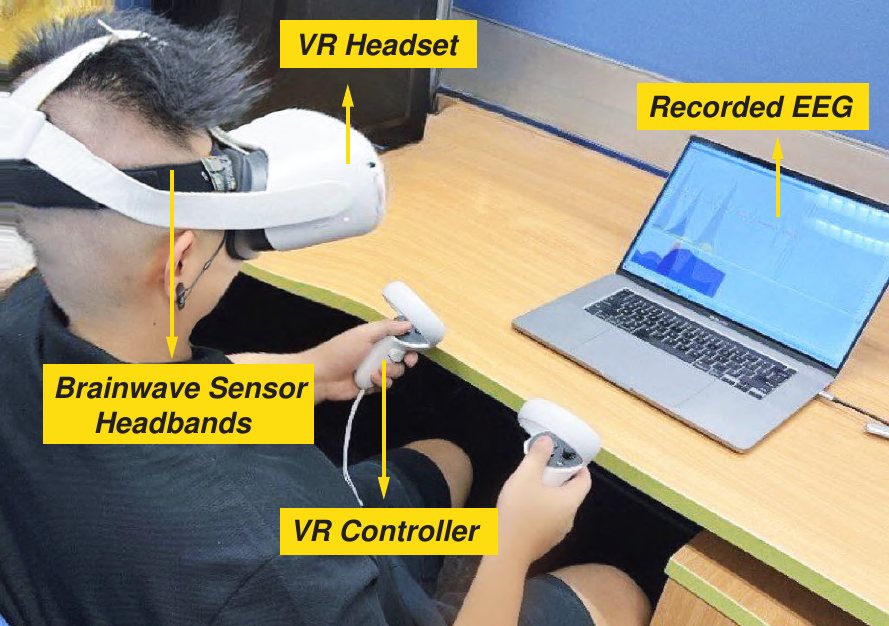}
         \vspace{-0.15in}
         \caption{Participant setup.}
         \label{fig:participant_setup}
    \end{subfigure}
    \begin{subfigure}[b]{.495\linewidth}
         \centering
         \includegraphics[width=\linewidth]{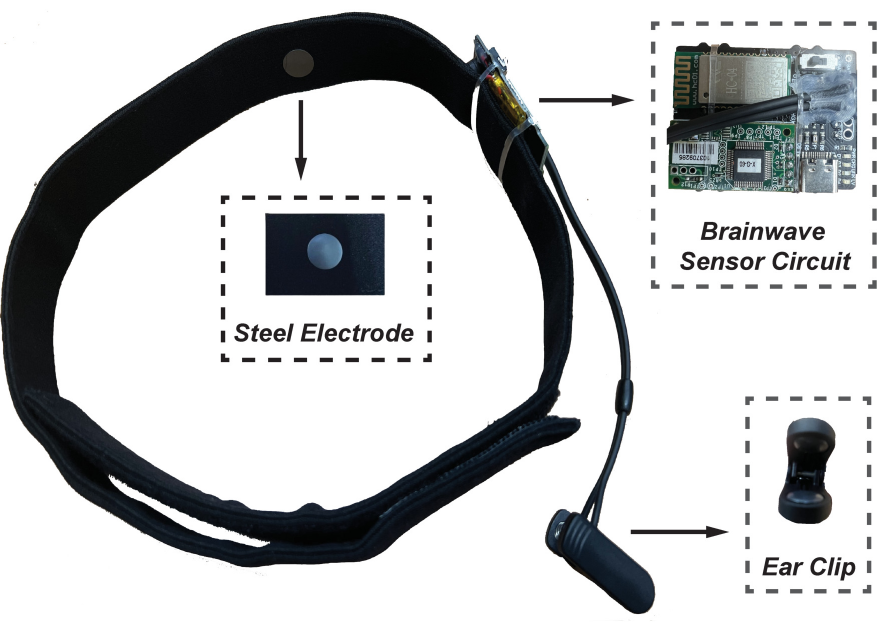}
         \vspace{-0.15in}
         \caption{Brainwave sensor headband.}
         \label{fig:brainwave_sensor}
    \end{subfigure}
    \vspace{-0.2in}
    \caption{Experiment setup for data collection.}
    \vspace{-0.2in}
    \label{fig:experimental_setup}
\end{figure}

To obtain real brain EEG signals from participants as groundtruth data for training reconstruction models, we utilize ThinkGear brainwave sensor headbands~\cite{thinkgearbrainsensor, mridha2021brain} as shown in \autoref{fig:brainwave_sensor}, which consist of a steel electrode to detect the EEG signals, an ear clip to obtain the reference signals and mitigate interference, and a TGAM board with a Bluetooth module to receive and transmit real-time EEG signals to the paired Android smartphone app.
Note that both the collected acceleration and EEG data samples from VR headsets' motion sensors are processed on a desktop computer.
 cGAN and transformer-based activity recognition models were trained on a NVIDIA RTX A6000 GPU.
\looseness=-1

\paragraph{Data Collection Process} 
As shown in \autoref{fig:participant_setup}, all participants were informed that motion sensor data from VR headsets and EEG signals from the sensors in the headband would be recorded.
Specifically, we recruit $25$ participants and evaluate \sysname by collecting data samples under the aforementioned three levels of attack surfaces\footnote{To support the Open Science policy, we make all relevant source code and supporting scripts publicly available on the Zenodo platform via the following permanent access: \url{https://doi.org/10.5281/zenodo.18957102}}.
\textit{(1) UI-level Data Collection:} In this scenario, each participant is required to open $50$ websites in default browsers, launch $50$ VR apps, and watch $50$ streaming videos within immersive VR apps such as \textsc{Netflix}. Specifically, the actions of opening websites and apps are repeated for $100$ times to collect sufficient data from both the VR's inbuilt motion sensors and the EEG headband. 
In particular, we collect data samples in the first minute of performing the above activities to guarantee the distinctive patterns in the collected signals induced by visual stimuli.
The complete lists of websites, VR apps, and videos are listed in \autoref{tab:appendix_merged_list} in the Appendix.
\textit{(2) User-level Data Collection: } For validating the user-level identity recognition, we collect data samples from the $25$ participants when starting the VR devices to make their eyes receive the same stimulus and repeat the collection procedure for $100$ times. To extract user-level gaze information, we ask each participant to perform eight types of gaze movements (\ie, U: up, D: down, L: left, R: right, UL: up left, UR: up right, DL: down left, and DR: down right) and collect the motion and EEG data in the same procedure.
\textit{(3) Brain-level Data Collection: } In this scenario, we collect data samples to demonstrate the feasibility of leveraging the reconstructed brainwave to infer brain-level perceptions, \ie, perceptive images in the human brain.
In particular, we show each participant images from common $40$ classes exploited in prior visual-to-brain works (\eg, ~\cite{bai2023dreamdiffusion, lan2023seeing, kavasidis2017brain2image, tirupattur2018thoughtviz, davis2022brain}) while repeatedly collecting the corresponding motion sensor data from VR and EEG signals from the brainwave sensors.
Specifically, the data collected from the EEG headband are also utilized to fine-tune the generative models derived from the pre-trained Stable Diffusion models.
We note that the requirement of collecting paired training data in the \textit{offline phase} does not violate our threat model, as the adversary can independently recruit participants to build training datasets in advance, while only requiring unrestricted motion sensor data from the victim during the \textit{online deployment} phase.


\begin{figure*}[t]
    \minipage{0.323\textwidth}%
    \centering
      \includegraphics[width=\linewidth]{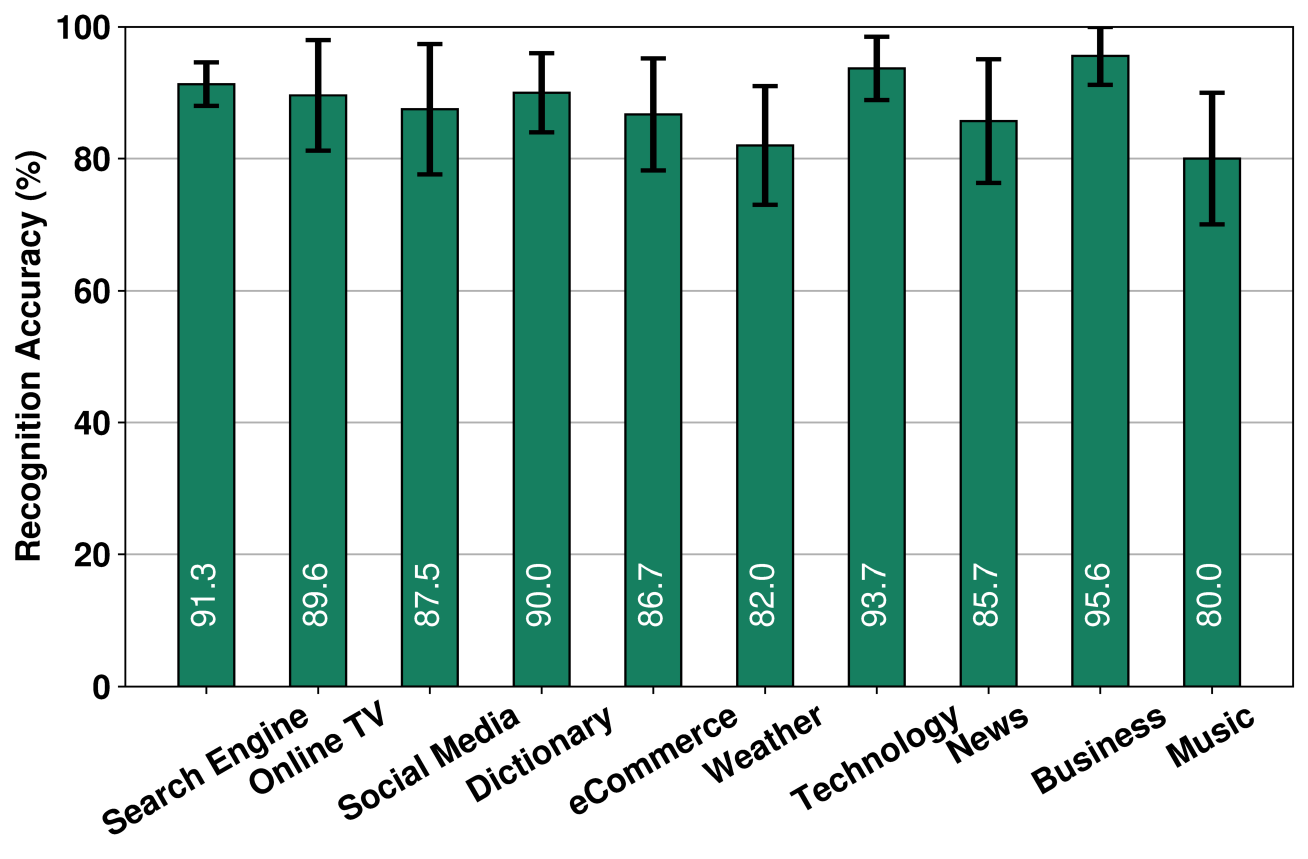}
      \vspace{-0.25in}
      \caption{UI-level website fingerprinting results with $50$ websites from $10$ categories.}
      \label{fig:overall_website_fingerprinting_results}
    \endminipage\hfill
    \minipage{0.323\textwidth}
    \centering
      \includegraphics[width=\linewidth]{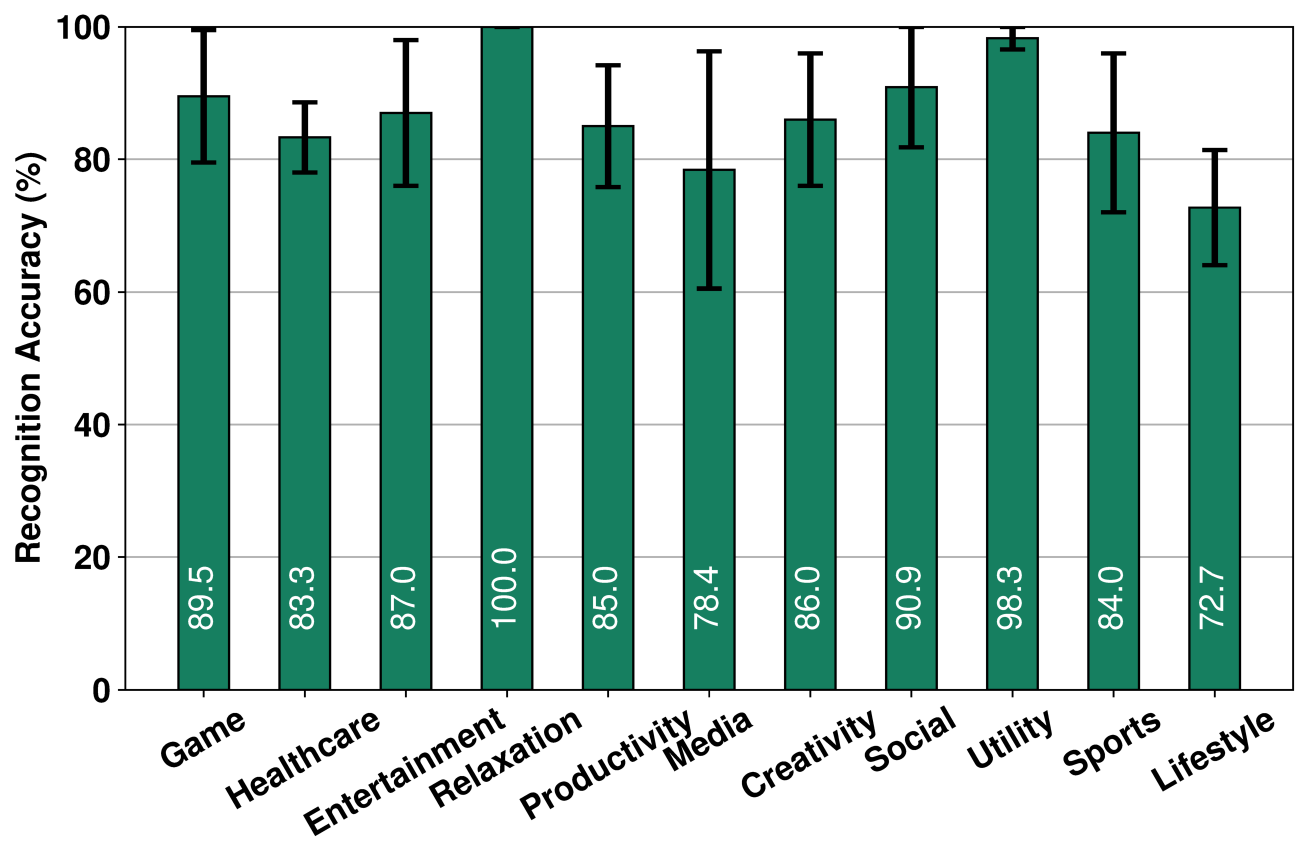}
      \vspace{-0.25in}
      \caption{UI-level VR app fingerprinting results with $50$ VR apps from $11$ categories.}
      \label{fig:overall_app_fingerprinting_results}
    \endminipage\hfill
    \minipage{0.323\textwidth}
    \centering
      \includegraphics[width=\linewidth]{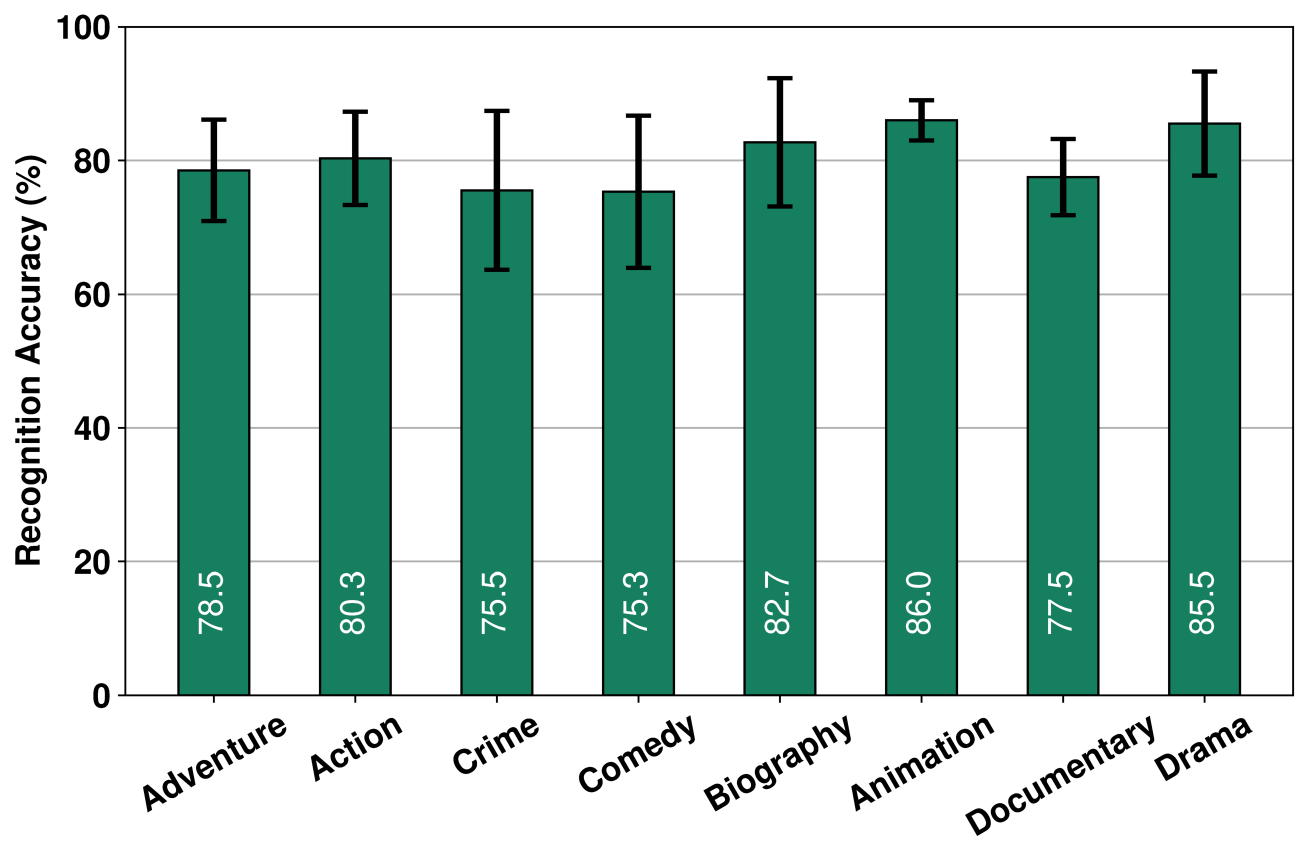}
      \vspace{-0.25in}
      \caption{UI-level Netflix video fingerprinting results with $50$ videos from $8$ categories.}
      \label{fig:overall_video_fingerprinting_results}
    \endminipage\hfill
    \vspace{0.05in}
    \minipage{0.3\textwidth}%
    \centering
      \includegraphics[width=\linewidth]{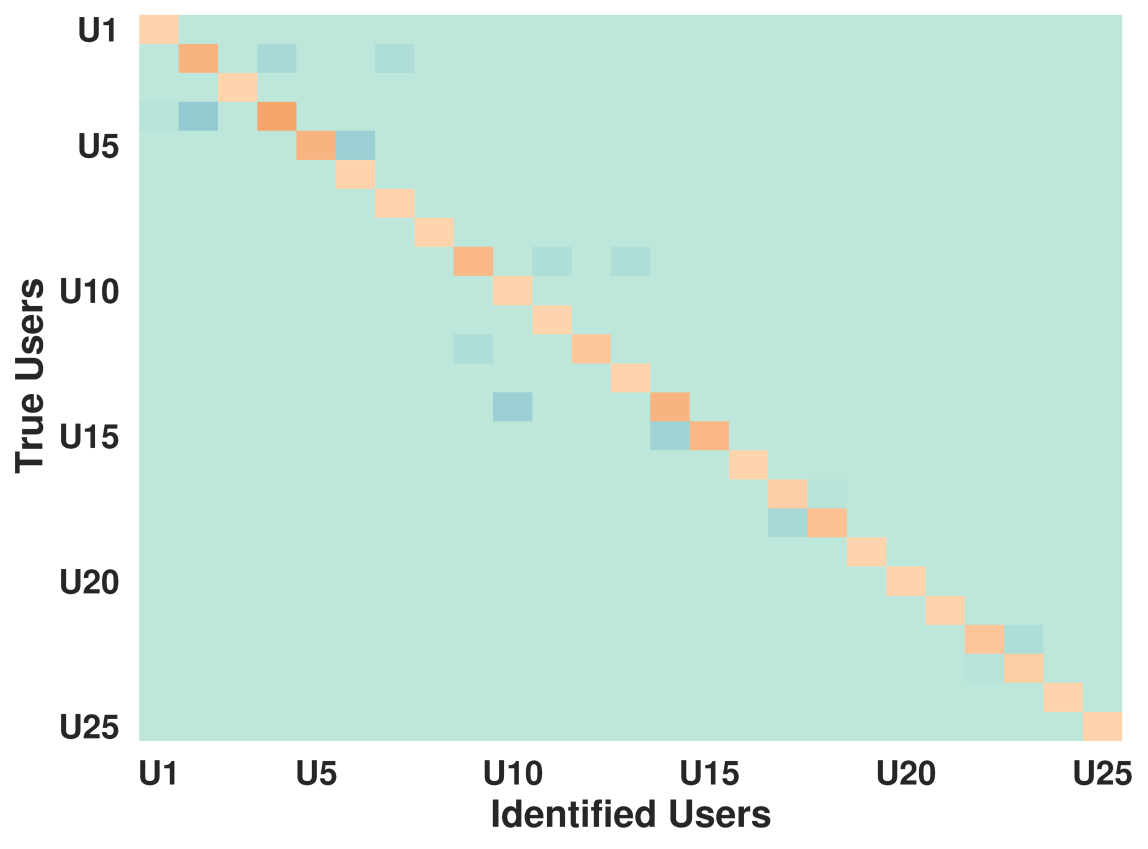}
      \vspace{-0.2in}
      \caption{User-level user de-anonymization. $U_{n}$: the $n$th participants.}
      \label{fig:overall_user_reidentification_results}
    \endminipage\hfill
    \minipage{0.3\textwidth}
    \centering
      \includegraphics[width=\linewidth]{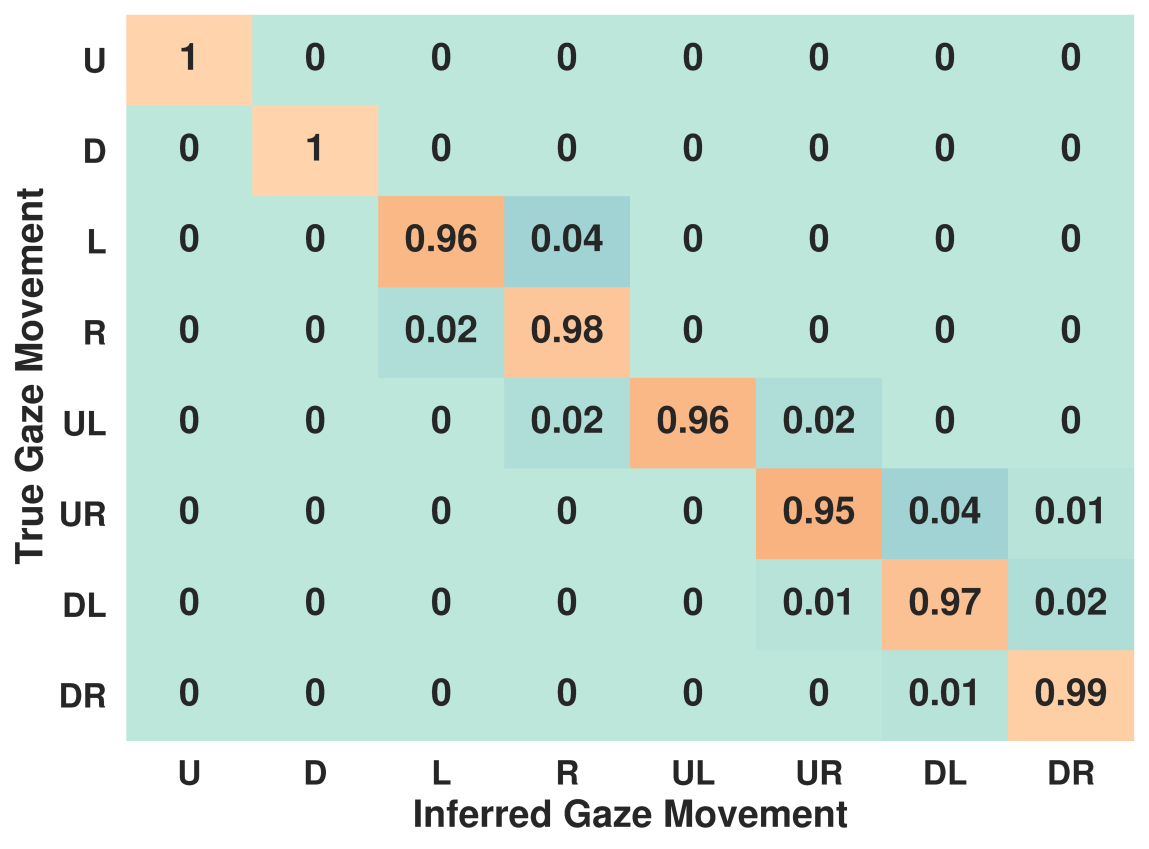}
      \vspace{-0.2in}
      \caption{User-level gaze movement recognition with different directions.}
      \label{fig:overall_gaze_movement_results}
    \endminipage\hfill
    \minipage{0.35\textwidth}
    \centering
      \includegraphics[width=\linewidth]{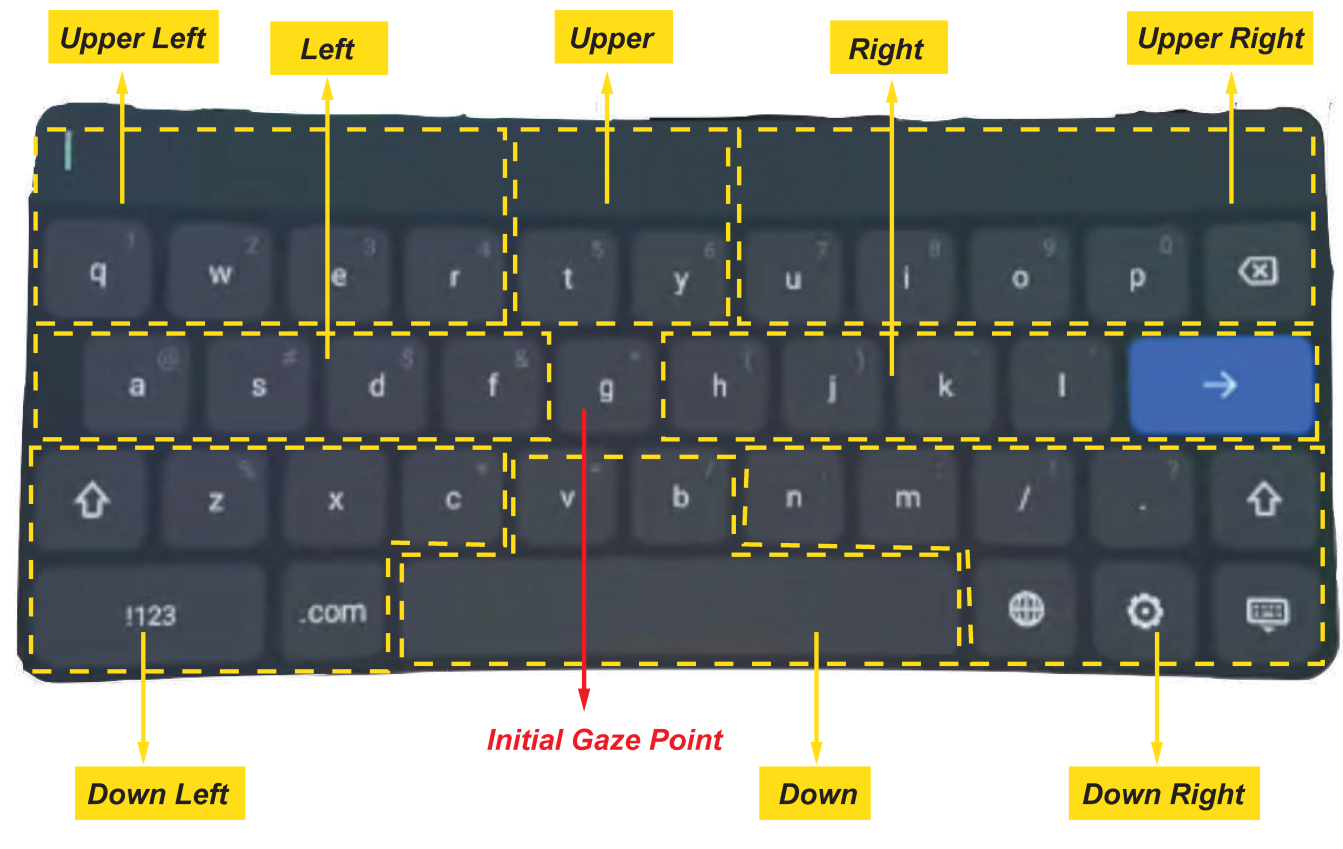}
      \vspace{-0.22in}
      \caption{Gaze-based keystroke inference by dividing virtual keyboard to eight zones.}
      \label{fig:keyboard_gaze_layout}
    \endminipage
    \vspace{-0.2in}
\end{figure*}

\paragraph{Evaluation Metrics} We use two metrics to quantify the effectiveness of \sysname in reconstructing different levels of user privacy within virtual scenes, including UI-level website/app/video fingerprinting, user-level identification de-anonymization and gaze movement recognition, as well as brain-level perceptive images inference.
\textit{(i) Accuracy:} We select accuracy as the metric for different levels of user privacy recognition, which is defined as the ratio of classes that are correctly identified.
\textit{(ii) Confusion matrix:} We evaluate the performance in recognizing user identities and gaze movements with confusion matrices, which visualize the actual target values alongside those predicted by the trained models.
In practice, we divide each collected data set into a training set and a testing set with a ratio of $8:2$. Training sets are used to train recognition models, and testing sets are used for performance evaluation.

\subsection{Evaluation of UI-level Privacy Inference}
\label{subsec:evaluation_ui_level_privacy_inference}

\paragraph{Website Fingerprinting Results} \autoref{fig:overall_website_fingerprinting_results} shows the evaluation results of \sysname in recognizing the $50$ different browsing websites, where it achieves an overall $89.3\%$ accuracy rate.
Among the evaluated websites, we observed that websites such as \textsc{twitch.tv} ($100\%$), \textsc{ticktock.com} ($100\%$), and \textsc{cnn.com} ($98\%$) are more recognizable when being browsed because these websites automatically load video streaming or many high-resolution images, leading to strong visual stimuli for the eyes that cause distinctive pupillary response patterns in reconstructed EEG signals.
On the contrary, \sysname presents an accuracy of only $70\%$ and $77\%$ in recognizing websites such as \textsc{google.com} and \textsc{duckduckgo.com},
which are widely used search engines with only a simple search bar and static page layout to render. Hence, these websites cause weak pupillary responses and present similar patterns that are easily misclassified.
\looseness=-1

\paragraph{VR App Fingerprinting Results} Furthermore, \autoref{fig:overall_app_fingerprinting_results} demonstrates that \sysname achieves an overall $85.8\%$ accuracy in fingerprinting $50$ VR apps from different categories.
\sysname achieves the highest performance ($100\%$) in fingerprinting VR apps from the game category (\eg, \textsc{Population: ONE} and \textsc{Cards \& Tankards}) and the media and streaming category (\eg, \textsc{Netflix} and \textsc{Pluto TV}), as these apps involve dynamic animations in the launch stage, resulting in strong visual stimuli to the user's eyes and causing drastic fluctuations in the reconstructed EEG signals.
However, \sysname performs worse in the recognition of apps that adopt the default app launch interface, \ie, a static white background with a simple app logo, such as social apps like \textsc{Multiverse} ($63\%$) and lifestyle apps such as \textsc{Maloka} ($51\%$),
whose reconstructed EEG patterns are plain due to limited visual stimuli.
\looseness=-1

\begin{figure*}[t]
    \minipage{0.33\textwidth}%
    \centering
      \includegraphics[width=\linewidth]{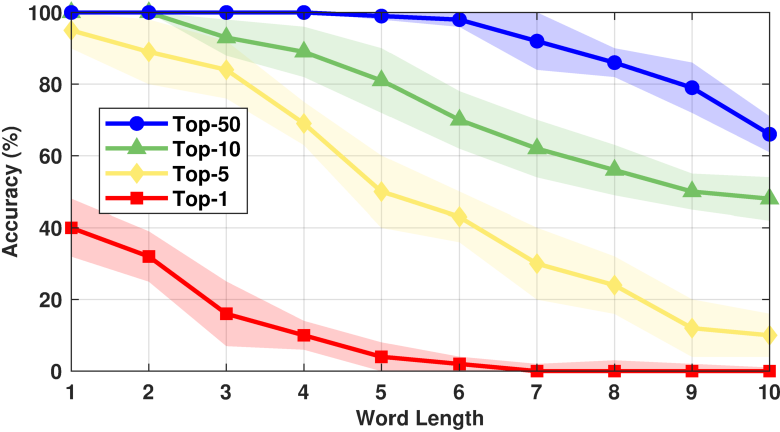}
      \vspace{-0.15in}
      \caption{Gaze-based keystroke inference results under top-1, 5, 10, 50 settings.}
      \label{fig:gaze_keystroke_inference}
    \endminipage\hfill
    \minipage{0.65\textwidth}
    \centering
      \includegraphics[width=\linewidth]{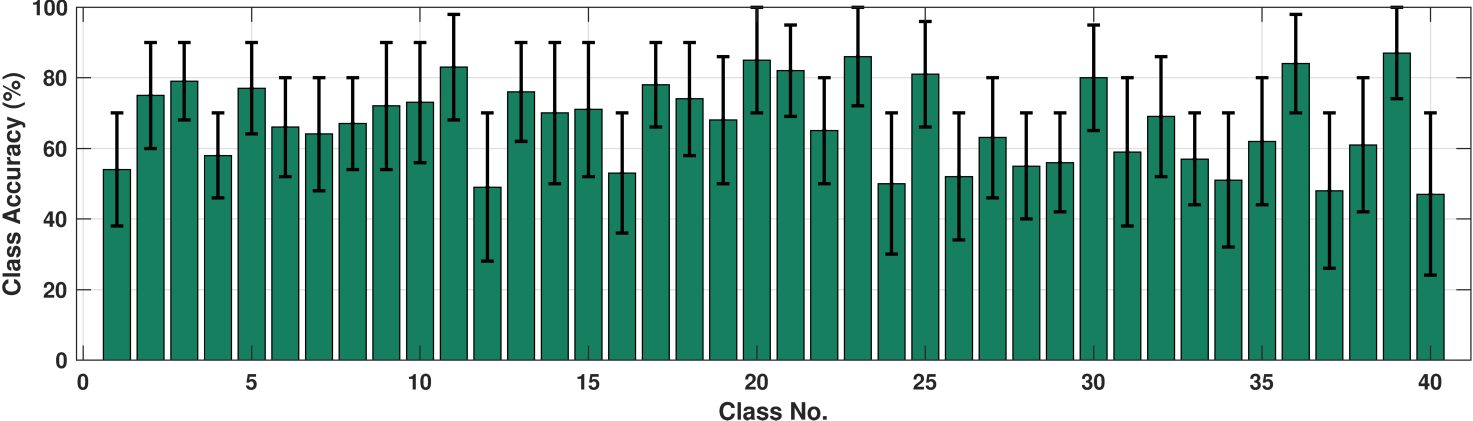}
      \vspace{-0.15in}
      \caption{Brain-level visual perception inference, including the image perceptions of $40$ classes in the ImageNet dataset~\cite{deng2009imagenet} of Stable Diffusion.}
      \label{fig:brain_activity_results}
    \endminipage
    \vspace{-0.2in}
\end{figure*}

\paragraph{Netflix Streaming Video Fingerprinting Results} Prior studies (\eg, \cite{nguyen2024penetration, gu2019traffic, enev2011televisions}) highlight the importance of investigating video fingerprinting that may leak user privacy, such as the political or religious interests of certain organizations or regions. Furthermore, in \autoref{fig:overall_video_fingerprinting_results}, we also present the evaluation results of fingerprinting $50$ streaming apps within \textsc{Netflix}.
In particular, \sysname achieves an average accuracy of $80.0\%$ in video fingerprinting, showcasing the highest performance when the VR user is watching streaming videos from the action category, like \textsc{Bandidos} ($95\%$) and \textsc{The Wages of Fear} ($93\%$), as they contain more dynamic content of activities and quick scene switching, making intense visual stimuli for the human eyes and strong brainwave responses. In contrast, streaming videos such as \textsc{Mea Culpa} ($57\%$) and \textsc{A Killer Paradox} ($54\%$) exhibit the worst performance because these two videos contain a lot of bland characters dialog and dark scenes, such that they cannot provide strong stimulation to the human eyes.
\looseness=-1

\subsection{Evaluation of User-level Identity Recognition and Gaze Movement Extraction}
\label{subsec:evaluation_user_level_identity_recognition_gaze_movement}

\paragraph{User Re-identification Results} User identity in virtual environments is highly sensitive to privacy concerns, which becomes susceptible to identity leakage caused by de-anonymization attacks~\cite{meng2023anonymization}.
Hence, we evaluate \sysname's performance in recognizing the $25$ VR participants' identities (denoted as $U_{1}$--$U_{25}$) by collecting the eye-blink clips to reconstruct the corresponding brain EEG signals.
\autoref{fig:overall_user_reidentification_results} shows the evaluation results of \sysname in user-level identity re-identification, where it achieves an overall $98.4\%$ accuracy rate.
Specifically, we observe that \sysname performs better in user-level re-identification than in UI-level privacy inference because eye blink results in apparent patterns that contain distinctive biometric characteristics due to the varying facial architecture of the participants.
The results also highlight \sysname's potential as a novel user authentication approach, leveraging biometric EEG signals reconstructed by the unrestricted motion sensors within the VR headset.
\looseness=-1

\paragraph{Gaze Movement Recognition Results} As illustrated in \autoref{subsec:experimental_setup}, we divide gaze movements into eight different directions to validate \sysname's ability to extract gaze, including upper (U), down (D), left (L), right (R), upper left (UL), upper right (UR), down left (DL), and down right (DR).
Overall, \autoref{fig:overall_gaze_movement_results} depicts that \sysname achieves $97.6\%$ accuracy in recognizing the eight directions of gaze movement.
Since gaze movements have been shown to reveal sensitive information (\eg, \cite{wang2019your, chen2018eyetell, abdrabou2022your}), such as UI-level keystrokes, 
we further investigate whether \sysname can also infer keystrokes on virtual keyboards from gaze data.
Specifically, we designate the key "G" as the focal point and divide the virtual keyboard, as shown in \autoref{fig:keyboard_gaze_layout}, into eight zones corresponding to the directions of gaze movement. This approach stems from the observation that eyes actively search for keys during the typing process. By analyzing the extracted gaze movements, we are then able to generate the top-k candidate combinations (\eg, $k=1, 5, 10, 50$) for each sequence of keys and subsequently infer the virtual keystrokes.
\looseness=-1

\autoref{fig:gaze_keystroke_inference} shows the results of \sysname in inferring $100$ words with lengths of one to ten using the extracted gaze movements.
The words are randomly selected from the Cambridge English vocabulary list~\cite{cambridgeenglish}.
We can observe that \sysname achieves an average $10.4\%$ top-1 accuracy, $50.6\%$ top-5 accuracy, $74.9\%$ top-10 accuracy, and $92.0\%$ top-50 accuracy, respectively.
Despite previous VR keystroke inference attacks (\eg, \cite{wu2023privacy, luo2022holologger, luo2024eavesdropping, zhang2023s}) that present higher recognition accuracy, our method, \sysname, introduces a novel orthogonal approach to realize UI-level virtual keystroke inference by reconstructing the brain EEG signals.
Note that the virtual keystroke inference accuracy of \sysname could be further enhanced if incorporated with previous state-of-the-art gaze analysis methods~\cite{wang2019your, chen2018eyetell, abdrabou2022your} and we only investigate its feasibility in this study.
\looseness=-1

\begin{table*}[t]
\caption{Robustness evaluation of \sysname under different scenarios.}
\label{tab:robust_evaluation_different_scenarios}
\scriptsize
\setlength{\tabcolsep}{2.7pt}
\renewcommand{\arraystretch}{1.0}
\begin{tabular}{ccccccccccccccccc}
\toprule
\multirow{2}{*}{\textbf{\begin{tabular}[c]{@{}c@{}}User\\ Privacy\end{tabular}}} & \multicolumn{4}{c}{\textbf{VR Headsets}} & \multicolumn{4}{c}{\textbf{Unseen Users}} & \multicolumn{4}{c}{\textbf{Sampling Rates}} & \multicolumn{4}{c}{\textbf{Light Intensities}} \\ \cmidrule(lr){2-5}\cmidrule(lr){6-9}\cmidrule(lr){10-13}\cmidrule(lr){14-17} 
 & Oculus 2 & Oculus & PICO 4 & HTC Vive & User$_{26}$ & User$_{27}$ & User$_{28}$ & User$_{29}$ & 100 Hz & 250 Hz & 500 Hz & 1000 Hz & 25\% & 50\% & 75\% & 100\% \\ \midrule
\multicolumn{1}{c|}{Website Fingerprinting (UI)} & \multicolumn{1}{c|}{89.3} & \multicolumn{1}{c|}{87.3} & \multicolumn{1}{c|}{88.7} & \multicolumn{1}{c|}{86.5} & \multicolumn{1}{c|}{87.2} & \multicolumn{1}{c|}{84.2} & \multicolumn{1}{c|}{88.6} & \multicolumn{1}{c|}{83.0} & \multicolumn{1}{c|}{70.5} & \multicolumn{1}{c|}{82.0} & \multicolumn{1}{c|}{89.3} & \multicolumn{1}{c|}{90.0} & \multicolumn{1}{c|}{75.2} & \multicolumn{1}{c|}{86.5} & \multicolumn{1}{c|}{89.3} & 79.3 \\ \hline
\multicolumn{1}{c|}{VR App Fingerprinting (UI)} & \multicolumn{1}{c|}{85.8} & \multicolumn{1}{c|}{83.6} & \multicolumn{1}{c|}{84.3} & \multicolumn{1}{c|}{82.0} & \multicolumn{1}{c|}{83.5} & \multicolumn{1}{c|}{81.0} & \multicolumn{1}{c|}{84.0} & \multicolumn{1}{c|}{79.1} & \multicolumn{1}{c|}{65.6} & \multicolumn{1}{c|}{80.5} & \multicolumn{1}{c|}{85.8} & \multicolumn{1}{c|}{86.1} & \multicolumn{1}{c|}{69.6} & \multicolumn{1}{c|}{82.7} & \multicolumn{1}{c|}{85.8} & 74.5 \\ \hline
\multicolumn{1}{c|}{Video Fingerprinting (UI)} & \multicolumn{1}{c|}{80.0} & \multicolumn{1}{c|}{76.0} & \multicolumn{1}{c|}{76.7} & \multicolumn{1}{c|}{75.5} & \multicolumn{1}{c|}{75.7} & \multicolumn{1}{c|}{72.0} & \multicolumn{1}{c|}{78.7} & \multicolumn{1}{c|}{70.5} & \multicolumn{1}{c|}{58.8} & \multicolumn{1}{c|}{73.7} & \multicolumn{1}{c|}{80.0} & \multicolumn{1}{c|}{80.5} & \multicolumn{1}{c|}{64.4} & \multicolumn{1}{c|}{78.7} & \multicolumn{1}{c|}{80.0} & 70.3 \\ \hline
\multicolumn{1}{c|}{User Re-identification (User)} & \multicolumn{1}{c|}{98.4} & \multicolumn{1}{c|}{92.9} & \multicolumn{1}{c|}{93.2} & \multicolumn{1}{c|}{92.8} & \multicolumn{1}{c|}{--} & \multicolumn{1}{c|}{--} & \multicolumn{1}{c|}{--} & \multicolumn{1}{c|}{--} & \multicolumn{1}{c|}{77.5} & \multicolumn{1}{c|}{91.7} & \multicolumn{1}{c|}{98.4} & \multicolumn{1}{c|}{98.8} & \multicolumn{1}{c|}{77.5} & \multicolumn{1}{c|}{95.5} & \multicolumn{1}{c|}{98.4} & 88.6 \\ \hline
\multicolumn{1}{c|}{Gaze Recognition (User)} & \multicolumn{1}{c|}{97.6} & \multicolumn{1}{c|}{90.0} & \multicolumn{1}{c|}{90.3} & \multicolumn{1}{c|}{89.6} & \multicolumn{1}{c|}{90.1} & \multicolumn{1}{c|}{86.6} & \multicolumn{1}{c|}{94.2} & \multicolumn{1}{c|}{82.3} & \multicolumn{1}{c|}{74.7} & \multicolumn{1}{c|}{89.6} & \multicolumn{1}{c|}{97.6} & \multicolumn{1}{c|}{98.3} & \multicolumn{1}{c|}{75.0} & \multicolumn{1}{c|}{93.4} & \multicolumn{1}{c|}{97.6} & 86.5 \\ \hline
\multicolumn{1}{c|}{EEG-based Perception (Brain)} & \multicolumn{1}{c|}{67.2} & \multicolumn{1}{c|}{66.6} & \multicolumn{1}{c|}{52.5} & \multicolumn{1}{c|}{52.0} & \multicolumn{1}{c|}{56.5} & \multicolumn{1}{c|}{50.8} & \multicolumn{1}{c|}{62.3} & \multicolumn{1}{c|}{48.4} & \multicolumn{1}{c|}{39.6} & \multicolumn{1}{c|}{63.5} & \multicolumn{1}{c|}{67.2} & \multicolumn{1}{c|}{68.0} & \multicolumn{1}{c|}{46.6} & \multicolumn{1}{c|}{62.0} & \multicolumn{1}{c|}{67.2} & 58.8 \\ \bottomrule
\end{tabular}
\vspace{-0.05in}
\end{table*}
\begin{table*}[t]
\caption{\minor{Ablation study results: \textit{Raw Acc.} (raw accelerometer) and \textit{Recon.\ EEG} (reconstructed EEG) under \textit{random split} and \textit{session-disjoint} split. $\Delta_{\text{EEG}}$: Improvement from EEG reconstruction under each split. $\Delta_{\text{SD}}$: Drop from session-disjoint evaluation.}}
\label{tab:ablation}
\scriptsize
\setlength{\tabcolsep}{9.5pt}
\renewcommand{\arraystretch}{1.0}
\begin{tabular}{ccccccccc}
\toprule
\multirow{2}{*}{\textbf{\begin{tabular}[c]{@{}c@{}}Multi-level User\\ Privacy\end{tabular}}} & \multicolumn{3}{c}{\textbf{Random Split}} & \multicolumn{3}{c}{\textbf{Session-Disjoint}} & \multicolumn{2}{c}{\textbf{$\Delta_{\text{SD}}$}} \\ \cmidrule(lr){2-4}\cmidrule(lr){5-7}\cmidrule(lr){8-9}
 & Raw Acc. & Recon.\ EEG & $\Delta_{\text{EEG}}$ & Raw Acc. & Recon.\ EEG & $\Delta_{\text{EEG}}$ & Raw Acc. & Recon.\ EEG \\ \midrule
\multicolumn{1}{c|}{Website Fingerprinting (UI)} & \multicolumn{1}{c|}{78.5} & \multicolumn{1}{c|}{89.3} & \multicolumn{1}{l|}{\up~+10.8} & \multicolumn{1}{c|}{74.6} & \multicolumn{1}{c|}{85.7} & \multicolumn{1}{l|}{\up~+11.1} & \multicolumn{1}{c|}{\down~$-$3.9} & \down~$-$3.6 \\ \hline
\multicolumn{1}{c|}{VR App Fingerprinting (UI)} & \multicolumn{1}{c|}{73.2} & \multicolumn{1}{c|}{85.8} & \multicolumn{1}{l|}{\up~+12.6} & \multicolumn{1}{c|}{69.1} & \multicolumn{1}{c|}{81.4} & \multicolumn{1}{l|}{\up~+12.3} & \multicolumn{1}{c|}{\down~$-$4.1} & \down~$-$4.4 \\ \hline
\multicolumn{1}{c|}{Video Fingerprinting (UI)} & \multicolumn{1}{c|}{67.8} & \multicolumn{1}{c|}{80.0} & \multicolumn{1}{l|}{\up~+12.2} & \multicolumn{1}{c|}{63.5} & \multicolumn{1}{c|}{75.8} & \multicolumn{1}{l|}{\up~+12.3} & \multicolumn{1}{c|}{\down~$-$4.3} & \down~$-$4.2 \\ \hline
\multicolumn{1}{c|}{User Re-identification (User)} & \multicolumn{1}{c|}{89.6} & \multicolumn{1}{c|}{98.4} & \multicolumn{1}{l|}{\up~+8.8} & \multicolumn{1}{c|}{84.8} & \multicolumn{1}{c|}{94.0} & \multicolumn{1}{l|}{\up~+9.2} & \multicolumn{1}{c|}{\down~$-$4.8} & \down~$-$4.4 \\ \hline
\multicolumn{1}{c|}{Gaze Recognition (User)} & \multicolumn{1}{c|}{85.3} & \multicolumn{1}{c|}{97.6} & \multicolumn{1}{l|}{\up~+12.3} & \multicolumn{1}{c|}{80.7} & \multicolumn{1}{c|}{93.5} & \multicolumn{1}{l|}{\up~+12.8} & \multicolumn{1}{c|}{\down~$-$4.6} & \down~$-$4.1 \\ \hline
\multicolumn{1}{c|}{EEG-based Perception (Brain)} & \multicolumn{1}{c|}{41.0} & \multicolumn{1}{c|}{67.2} & \multicolumn{1}{l|}{\up~+26.2} & \multicolumn{1}{c|}{35.8} & \multicolumn{1}{c|}{59.6} & \multicolumn{1}{l|}{\up~+23.8} & \multicolumn{1}{c|}{\down~$-$5.2} & \down~$-$7.6 \\ \bottomrule
\end{tabular}
\vspace{-0.2in}
\end{table*}

\subsection{Evaluation of Brain-level Image Perception}
\label{subsec:evaluation_brain_level_activity_recognition}

In this section, we further evaluate the performance of \sysname in recognizing brain-level visual perceptions.
Due to the non-deterministic characteristics of the Stable Diffusion model, the output perceptual images from the fine-tuned model vary, whereas images in the same class often present similar patterns~\cite{bai2023dreamdiffusion}.
Specifically, we utilize \textit{Class Accuracy} (CA) as the evaluation metric to assess \sysname's performance.
That is, if the output image and the ground truth image belong to the same class (\eg, ``Cat'', ``Dog'', \etc), we consider it a successful trial of recognizing brain-level activity.
\looseness=-1

In \autoref{fig:brain_activity_results}, the class accuracy of perceptive image inference across the aforementioned 40 classes is presented, demonstrating that \sysname achieves an overall class accuracy rate of $67.2\%$. In particular, accuracy rates vary significantly between different classes due to image heterogeneity within each class. For example, \sysname achieves high accuracy rates of $79\%$ for the ``Butterfly'' class (No. 3) and $83\%$ for the ``Kayak'' class (No. 11), mainly because most of the images in these classes consist of similar and easily recognizable elements, such as green leaves and red kayaks floating on water surfaces, respectively. In contrast, classes with heterogeneous images, such as ``Cat'' ($54\%$) and ``Cellular telephone'' ($49\%$), demonstrate lower performance and limited channel capacity, attributed to the diverse variations within these categories. Despite these challenges, \sysname establishes itself as the pioneering work in reconstructing brain-level image perceptions from the inbuilt accelerometer in the VR headset, showcasing competitive performance.
\looseness=-1

\subsection{Ablation Study}
\label{subsec:evaluation_ablation_study}

To validate the necessity of EEG reconstruction and rule out potential evaluation artifacts, we conduct ablation studies on the Meta Oculus Quest 2 in two dimensions: \textit{(1) Input Type:} We compare raw accelerometer spectrograms against reconstructed EEG spectrograms, and \textit{(2) Evaluation Protocol:} We compare the standard random 8:2 split against a session-disjoint split, in which training and testing data are drawn from separate collection sessions at least one week apart to control for session-specific artifacts, such as headset placement and user posture. 
\autoref{tab:ablation} presents the empirical results of all four combinations, which show that: (1) EEG reconstruction consistently improves performance, with the most significant gain at the brain level, \ie, +$26.2\%$ under random split, +$23.8\%$ under session-disjoint, which rules out the possibility that the model only predicts class labels from head-motion cues, (2) The session-disjoint protocol introduces only moderate drops, \ie, $3.6\%$--$4.4\%$ for UI-level, $4.1\%$--$4.4\%$ for user-level, and $7.6\%$ for brain-level, which confirms that \sysname captures genuine stimulus-response patterns rather than session-specific correlations, and (3) Even in the strictest setting combining raw accelerometer input with session-disjoint splits, \sysname achieves $74.6\%$ website fingerprinting and $35.8\%$ brain-level perception inference, both substantially above random baselines, \ie, $2.0\%$ and $2.5\%$, respectively.



\section{Robustness of \sysname}
\label{sec:evaluation_robustness}

\paragraph{Different VR Headsets} Given the diverse hardware designs of various VR headsets, such as differences in accelerometers and gyroscopes, the recorded acceleration signals may exhibit subtle variations in patterns.
To evaluate the robustness and extensibility of \sysname to other VR headsets, we carried out further experiments applying the Meta Oculus Quest 2 to data samples collected from three other VR devices, including the Meta Oculus Quest, PICO 4 All-in-One, and HTC Vive Pro.
The evaluation results of the three-level user privacy inference based on reconstructed brain EEG signals from these three VR headsets are depicted in \autoref{tab:robust_evaluation_different_scenarios}.
Notably, \sysname maintains high recognition accuracy in UI-level inference, achieving  $86.5\%$--$88.7\%$ in fingerprinting $50$ websites, $82.0\%$--$84.3\%$ in fingerprinting 50 VR apps, and $75.5\%$--$76.7\%$ in fingerprinting $50$ streaming videos on Netflix, respectively.
However, there is a slight decrease in \sysname's performance in user-level recognition, from $98.4\%$ to $92.9\%$, $93.2\%$, and $92.8\%$ in user re-identification and from $97.6\%$ to $90.0\%$, $90.3\%$, and $89.6\%$ in gaze movement extraction and keystroke inference. 
In particular, there is a significant decrease in \sysname's performance in brain-level perception recognition when directly transferred to PICO 4 and HTC Vive VR headsets, decreasing from $67.2\%$ to $52.5\%$, $52.0\%$, respectively.
This decrease can be attributed to the varying sensitivity of the inbuilt motion sensors in these VR headsets from different brands, leading to additional perturbations in the input acceleration signals and resulting in a performance decrease in the EEG reconstruction models.
Nevertheless, this issue can be mitigated by collecting more data samples from the new VR headset to fine-tune the pre-trained recognition models.

\paragraph{Different VR Users} Due to the inherent variability in biometric facial characteristics among different participants, pupillary responses to identical visual stimuli can exhibit subtle differences, introducing variations in reconstructed EEG signals.
To comprehensively understand the impact of participant diversity, we recruit four additional participants, denoted as $U_{26}$--$U_{29}$, and collect unseen data samples from them.
We then apply the pre-trained models to evaluate \sysname's robustness across these new users.
Specifically, the evaluation results shown in \autoref{tab:robust_evaluation_different_scenarios} reveal that \sysname's performance presents an average decrease of $4.4\%$ across various UI-level inference, \ie, $3.6\%$ website fingerprinting, $3.9\%$ app fingerprinting, and $5.8\%$ video fingerprinting. 
Likewise, we observe a decrease of $9.3\%$ in user-level gaze movement recognition and keystroke inference, along with a decrease of $12.7\%$ in brain-level perception recognition. However, \sysname maintains acceptable performance in inferring fine-grained user privacy, and the observed reduction in performance was predominantly due to user variability resulting from the limited datasets collected for model training.
Thus, addressing this limitation involves collecting data samples from more extensive and diverse participants.
\looseness=-1

\paragraph{Different Accelerometer Sampling Rates} In \autoref{subsec:experimental_setup}, we set the sampling rate of the inbuilt accelerometer to \SI{500}{\hertz} to capture acceleration signals to reconstruct brain EEG signals.
As different sampling rates encode user activity information with varying granularity, we conducted additional experiments to explore their impact by collecting data samples from the accelerometer at rates of \SI{100}{\hertz}, \SI{250}{\hertz}, \SI{500}{\hertz}, and \SI{1000}{\hertz} using the Meta Oculus Quest 2 VR headset, then assessed performance using pre-trained models at the default \SI{500}{\hertz} rate, respectively.
The evaluation results, depicted in \autoref{tab:robust_evaluation_different_scenarios}, reveal an increase in \sysname's performance with higher sampling rates.
When setting the sampling rate at \SI{100}{\hertz}, \sysname exhibited significantly lower performance compared to the default rate, achieving only $65.0\%$ in UI-level, $76.1\%$ in user-level, and $39.6\%$ in brain-level inference, respectively.
In particular, as sampling rates beyond \SI{500}{\hertz} offer marginal increments because of the limited frequency range of pupillary responses, we conclude that \SI{500}{\hertz} achieves a balance between \sysname's performance and stealthiness, considering the higher energy consumption associated with higher sampling frequencies, which may raise suspicions.
\looseness=-1

\paragraph{Different Light Intensities of the Virtual Scene} As shown in previous studies~\cite{shen2022pupilrec, lan2020gazegraph}, light intensity could impact the response time of the human eye. Since \sysname relies on pupillary responses to reconstruct brain EEG signals and infer user privacy, the light intensity of virtual scenes (\aka, brightness) becomes a significant factor influencing visual stimuli.
In our previous evaluation detailed in \autoref{sec:evaluation}, we maintain default settings with a brightness level set to $75\%$.
To study the impact of varying light intensities on the performance of \sysname, we adjusted the brightness levels to $25\%$, $50\%$, and $100\%$.
The evaluation results, presented in \autoref{tab:robust_evaluation_different_scenarios}, show the optimal performance of \sysname at the $50\%$ and $75\%$ brightness levels.
We also observe that \sysname's performance diminishes under low-light conditions ($25\%$ brightness) due to weak visual stimuli, which lead to less apparent pattern changes in the pupillary responses.
On the other hand, at \SI{100}{\percent} brightness, the increased light intensity decreases the eyes' sensitivity to content changes in the VR headset's head-mounted display, resulting in a decrease in pupillary responses and pattern distinctiveness associated with various user activities and lower recognition accuracy rates.
\looseness=-1

\section{Countermeasures}
\label{sec:discussion_countermeasures}

\paragraph{Permission-based and Privacy-aware Management} Similar to previous works (\eg, ~\cite{wu2023privacy, slocum2023going, zhang2023s, meteriz2022keylogging, luo2022holologger}), \sysname requires access to unrestricted motion sensors in the VR headset. Hence, a potential countermeasure to defend against privacy leakage is to design and implement a privacy-based and permission-based sensor management scheme in the VR platform, as proposed on Android platforms~\cite{han2017sendroid, li2021android}.
For instance, VR users would need to grant permission for VR apps or webpages to access specific sensors during both installation and runtime. Furthermore, to improve transparency and control over data usage, the sensor management scheme should be privacy-aware, providing comprehensive information on the types of sensors, their activation times, durations of operation, and data collection activities in both the foreground and background.
Using contextual information, the privacy-aware sensor management scheme could allow users to create and update customized access control policies for all embedded sensors, enabling them to limit data access for background apps. Additionally, the framework assesses the quality of sensor data flow to VR apps and allows users to customize specific data attributes (\eg, sampling frequency, resolution) for a specific VR app.

\begin{figure}[t]
    \begin{subfigure}[b]{.495\linewidth}
         \centering
         \includegraphics[width=\linewidth]{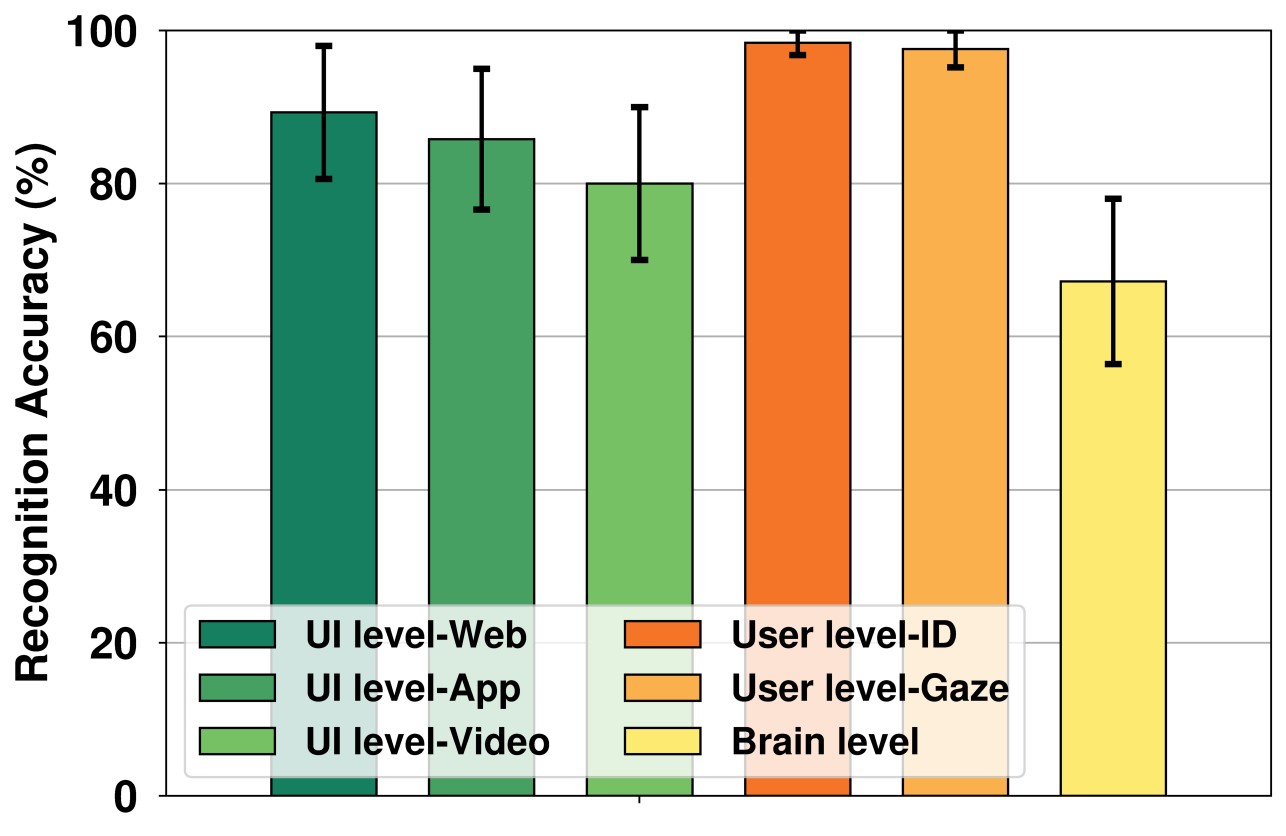}
         \vspace{-0.2in}
         \caption{Before applying defense.}
         \label{fig:before_defense}
    \end{subfigure}
    \begin{subfigure}[b]{.495\linewidth}
         \centering
         \includegraphics[width=\linewidth]{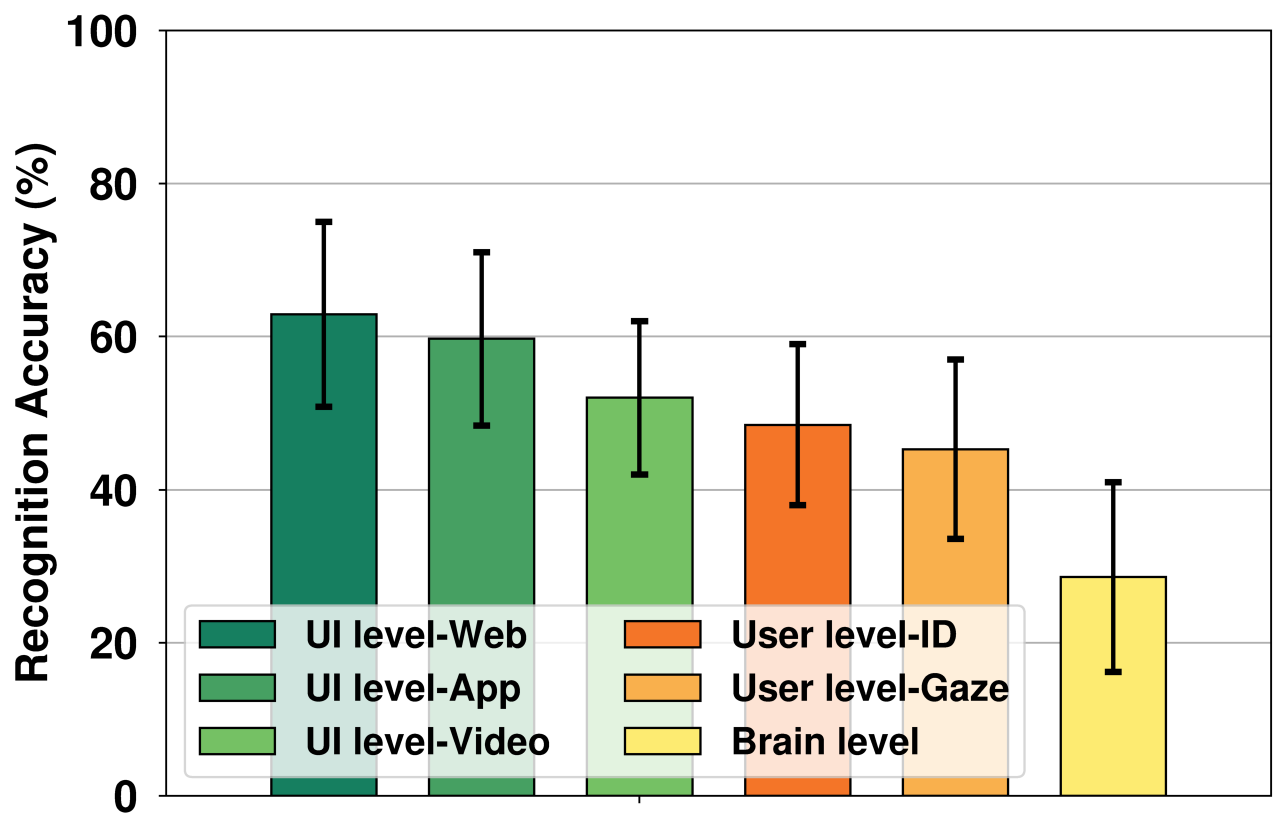}
         \vspace{-0.2in}
         \caption{After applying defense.}
         \label{fig:after_defense}
    \end{subfigure}
    \vspace{-0.15in}
    \caption{Empirical results of \sysname before and after applying sensor signal obfuscation defense method.}
    \vspace{-0.2in}
    \label{fig:empirical_results_defense}
\end{figure}

\paragraph{Sensor Signal Obfuscation} Another countermeasure is to apply a signal obfuscation method when accessing data from unrestricted motion sensors at a specific time.
For example, the VR system could add random noise (\eg, Gaussian white noise) to the motion sensor data when launching apps or opening websites, which could prevent the UI-level privacy inference from \sysname.
Besides, we could exploit data transformation and encryption methods~\cite{rastogi2010differentially, malekzadeh2020privacy, ni2023eavesdropping} to mask motion sensor data to hide sensitive information and retain data that are necessary to support the specific VR app to reduce recognition effectiveness and mitigate threats from user-level de-anonymization and gaze data leakage, as well as brain-level perceptions inference.
In practice, we designed an automatic, software-defined mechanism to apply Gaussian white noise to the recorded data from unrestricted motion sensors in Meta Oculus Quest 2 with a standard deviation of $0.1$ and evaluated \sysname's performance on UI-level, user-level, and brain-level privacy inferences.
\autoref{fig:empirical_results_defense} shows that the accuracy of the three levels of privacy inference decreases approximately $26.8\%$, $51.1\%$, and $38.6\%$, respectively.
However, obfuscating the signals from unrestricted motion sensors could also decrease the usability of VR headsets (\eg, long latency, slower response), which rely on these sensor readings to provide their real-time position and movement tracking ability.
In addition, recent studies~\cite{meteriz2022keylogging, wu2023privacy} have demonstrated the effectiveness of defending against inference attacks by randomly shuffling the layouts of some privacy-related UI, such as virtual keyboards, when entering sensitive information like accounts and passwords.
\looseness=-1

\section{Limitations}
\label{sec:limitations}

Although \sysname reveals both observable and unobservable user privacy leakage, our framework still has some limitations.
First, as a proof-of-concept study, we only evaluated one type of brain activity, \ie, validated its class-level perception inference across $40$ image categories shown in DreamDiffusion~\cite{bai2023dreamdiffusion}, and additional EEG-image pairs from other brain-level activity inference will require fine-tuning models or utilizing EEG-ImageNet~\cite{zhu2024eeg} (\eg, 80 image classes), or exploiting EEG-to-video backbones like EEG2Video~\cite{liu2024eeg2video} to expand both the scope and granularity of brain-level inference.
Second, it shows limited generalizability and performs well in relatively controlled scenarios such as browsing or light interaction, and the brain-level perception inference degrades across different headsets, which could be mitigated by cross-subject domain adaptation, \ie, domain-adversarial neural networks~\cite{liu2024eeg2video} and adaptive feature representation learning methods~\cite{chen2023cinematic}.
In addition, while the subtle signals of interest could potentially be overwhelmed by intensive movements, our body movement mitigation module (\autoref{subsec:motion_data_processing}) addresses this issue and improves the signal-to-noise ratio (SNR) of the acceleration signals from approximately $11$--\SI{21}{\deci\bel}. Moreover, prior studies have demonstrated even weaker physiological signals from VR motion sensors at comparable or lower SNR levels, such as heartbeats ($5$--\SI{15}{\deci\bel})~\cite{zhang2023facereader} and blood pressure-related vibrations ($5$--\SI{10}{\deci\bel})~\cite{ye2025bpsniff}, which further validate the feasibility of our approach.


\section{Related Works}
\label{sec:related_works}

\paragraph{Attacks on VR Devices} With the rise of the Metaverse, recent studies have investigated VR attacks to steal sensitive information, \eg, and keystrokes on the virtual keyboard.
Most of the previous work installs malicious websites or apps to acquire readings from unrestricted motion sensors inside the VR headset and trains specific DNN models to reveal virtual keyboard input (\eg, \cite{wu2023privacy, zhang2023s, slocum2023going, meteriz2022keylogging, luo2022holologger, nair2023unique, ling2019know, zhang2023facereader, lee2025eyes, khalili2024virtual, ni2024sensor}).
Face-Mic~\cite{shi2021face} and ImmerSpy~\cite{cayir2025speak} show the feasibility of speech eavesdropping using zero-permission motion data.
Recent studies also utilize cameras to record user videos while wearing the VR headset and infer keyboard input~\cite{gopal2023hidden, meteriz2022keylogging} from watching videos~\cite{nguyen2024penetration} in virtual scenes and other user's keystrokes in the surrounding real environment~\cite{yang2023towards}.
Moreover, attackers can exploit virtual avatars to de-anonymize user identities~\cite{meng2023anonymization} and infer real-world keyboard typing through the avatars' movements~\cite{yang2024can, su2024remote}.
VR-Spy~\cite{al2021vr}, Heimdall~\cite{luo2024eavesdropping} and VRecKey~\cite{ni2024non} present side-channel attacks using Wi-Fi signals, button click sounds, and leaked infrared signals from VR controllers to infer virtual keystrokes, respectively.
In addition, recent research efforts have shown that attackers could use motion sensor data to infer biometric features, such as respiration rates and heartbeats~\cite{zhang2023facereader}, gender and body fat ratio~\cite{zhang2023passive}, and blood pressure~\cite{ye2025bpsniff}.
Compared with previous work, our work, \sysname, shows the feasibility of brainwave reconstruction with unrestricted motion sensors in VR headsets, which could lead to various attack surfaces.
\looseness=-1

\paragraph{Brain Activity Inference via BCI} Most of the initial research on neural decoding and brain activity inference focuses on exploiting functional magnetic resonance imaging (fMRI) (\eg,
\cite{takagi2023high, lin2022mind}) or electroencephalography (ECoG)~\cite{costecalde2018long}.
Specifically, researchers scan the human brain with professional BCI facilities (\eg, fMRI scanner~\cite{snow2011bringing}) under different visual stimuli from real images, and then input the fMRI data into generative neural networks (\eg, GAN~\cite{fang2020reconstructing}, diffusion models~\cite{lu2023minddiffuser}) to uncover brain activities.
However, due to the difficulty of obtaining fMRI data, recent studies have demonstrated the feasibility of leveraging EEG signals extracted from human brains to decode brain activities, \ie, perceptive images (\eg, ~\cite{bai2023dreamdiffusion, kavasidis2017brain2image, davis2022brain}), which can be extracted using non-invasive and portable BCI devices (\eg, brainwave sensor headbands~\cite{bciheadband}).
In particular, a recent study~\cite{tarkhani2022enhancing} also explored the potential attack vectors from the side of the operating, system as well as various on-device models integrated with commercial BCI platforms.
Our work, \sysname, becomes the \textit{first} work to demonstrate the feasibility of bridging the gap between the Metaverse and BCI by utilizing unrestricted motion sensors in head-mounted VR headsets to reconstruct
brain EEG signals.


\section{Conclusion}
\label{sec:conclusion}

We propose \sysname, a novel system for reconstructing brain EEG signals by exploiting inbuilt unrestricted motion sensors of a VR headset.
Specifically, \sysname leverages the acceleration signals induced by pupillary responses under various visual stimuli to reconstruct EEG signals.
Then, the adversary can exploit the reconstructed brain EEG signals to infer observable UI-level user activities (\eg, browsing websites, launching apps, and streaming videos), user-level identity de-anonymization and gaze movement extraction, and unobservable brain-level image perception.
The extensive evaluation suggests that \sysname achieves high accuracy in recognizing the privacy of three levels of users within the virtual world and the human mind.
To our knowledge, \sysname is the first study to highlight unexplored avenues of unobservable activities of brain-level perception inference among all newly proposed VR-related research.
\looseness=-1

\section*{Ethics Considerations}
\label{subsec:ethical_considerations}

We take ethical considerations seriously, and this study was approved by the
Human Subjects Ethics Sub-Committee of City University of Hong Kong (No. HU-STA-00000169).
We recruited $25$ volunteers from university students and staff ($15$ males and $10$ females, with ages ranging from $18$ to $35$) for data collection in this study.
In particular, we observed that only $11$ participants have previous VR experience, while the other $14$ participants do not have knowledge of using VR devices.
Hence, we asked participants to perform different activities in VR devices for one hour to become familiar with VR interactions while wearing the headband with inbuilt EEG brainwave sensors before the official data collection.

Before the experiments, each participant must sign a written consent form that allows us to collect data on human behavior for evaluation.
During the experiments, participants could move slightly while sitting or standing, as they were casually playing with the VR device,
and we only used our own accounts on VR platforms to browse websites, run apps, play
VR games, and type keystrokes.
Considering the dizziness and motion sickness associated with VR usage, we require participants to rest for $1$--$10$ minutes before reporting their conditions after collecting data with different time durations in each trial.
Note that we spent one year on the data collection, and all collected data samples are securely stored on a encrypted local server to prevent
any form of privacy leakage pertaining to the volunteers, which is only accessible by the authorized principal instructor and the graduate students in this project.

\section*{Acknowledgment}

We sincerely appreciate our shepherd and all anonymous reviewers for their constructive feedback and invaluable comments.
This work was fully supported by the Hong Kong Research Grants Council (RGC) under Grants CityU 21219223, 11219524, 11219025, RFS2122-1S04, C1029-22G, C6015-23G, CRS\_HKUST601/24, DON\_RMG 9229170, and AC-202403-02-15.
%
Any opinions, findings, and conclusions in this paper are those of the authors and are not necessarily those of the supported organizations. 





\bibliographystyle{IEEEtran}
\bibliography{reference}
%




\onecolumn
\appendices

\section{Supplementary Figures and Tables}
\label{apdx:figures_tables}

\begin{figure}[H]
\centering
    \begin{subfigure}[b]{.325\textwidth}
         \centering
         \includegraphics[width=\linewidth]{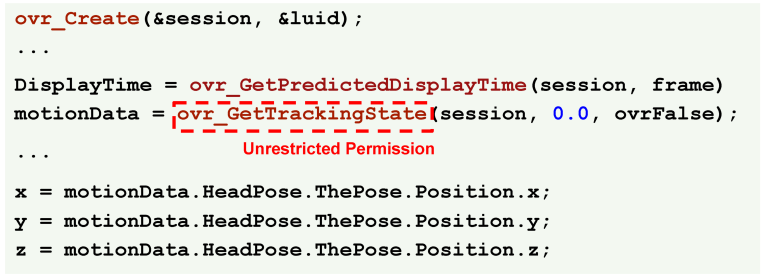}
         \vspace{-0.1in}
         \caption{Oculus Mobile SDK.}
         \label{fig:code_snippet_oculus}
    \end{subfigure}
    \hfill
    \begin{subfigure}[b]{.325\textwidth}
         \centering
         \includegraphics[width=\linewidth]{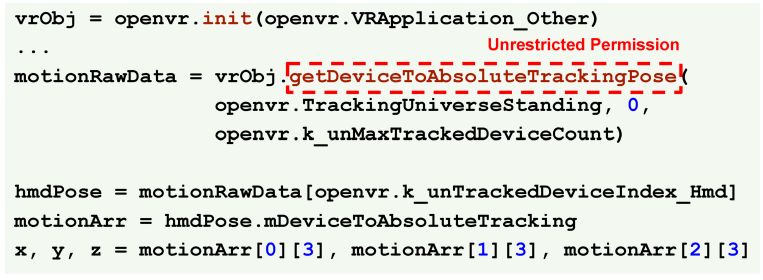}
         \vspace{-0.1in}
         \caption{OpenVR SDK.}
         \label{fig:code_snippet_openvr}
    \end{subfigure}
    \hfill
    \begin{subfigure}[b]{.325\textwidth}
         \centering
         \includegraphics[width=\linewidth]{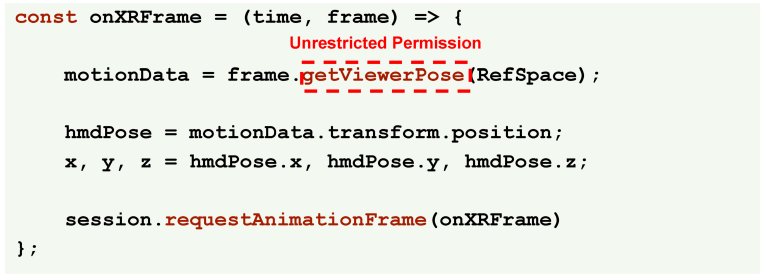}
         \vspace{-0.1in}
         \caption{WebXR Device API.}
         \label{fig:code_snippet_webxr}
    \end{subfigure}
    \caption{Code snippets depict the accessing built-in motion sensor data with unrestricted permission on three VR APIs/SDKs.}
    \label{fig:code_snippets}
    \vspace{-0.1in}
\end{figure}

\begin{table}[H]
\centering
\tiny
\setlength{\tabcolsep}{1.1pt}
\renewcommand{\arraystretch}{1.6}
\caption{Full list of $50$ VR apps, $50$ Netflix videos, and $50$ websites used in the evaluation (\autoref{subsec:experimental_setup}) as of May 2025.}
\label{tab:appendix_merged_list}
\resizebox{\textwidth}{!}{%
\begin{tabular}{@{}l l l l @{\hspace{4pt}} l l l l @{\hspace{4pt}} l l l l@{}}
\toprule
\multicolumn{4}{c}{\textbf{VR Apps}} & \multicolumn{4}{c}{\textbf{Netflix Videos}} & \multicolumn{4}{c}{\textbf{Websites}} \\
\cmidrule(lr){1-4}\cmidrule(lr){5-8}\cmidrule(lr){9-12}
\textbf{\#} & \textbf{App} & \textbf{Category} & \textbf{\# Ratings} &
\textbf{\#} & \textbf{Video} & \textbf{Category} & \textbf{\# Votes} &
\textbf{\#} & \textbf{Website} & \textbf{Category} & \textbf{Monthly Visits} \\
\midrule
$A_{1}$ & Population: One & Games & \appbar{14761}{14761} & $V_{1}$ & Society of the Snow & Adventure & \appbar{129537}{129537} & $W_{1}$ & google.com & Search engines & \webbar{162500000000}{162500000000} \\
$A_{2}$ & FitXR & Health \& Fitness & \appbar{14761}{9436} & $V_{2}$ & The Gentlemen & Action & \appbar{129537}{82777} & $W_{2}$ & youtube.com & Online TV & \webbar{162500000000}{106500000000} \\
$A_{3}$ & Gun Raiders & Games & \appbar{14761}{4919} & $V_{3}$ & 3 Body Problem & Adventure & \appbar{129537}{81461} & $W_{3}$ & facebook.com & Social media networks & \webbar{162500000000}{17500000000} \\
$A_{4}$ & Epic Roller Coasters & Games & \appbar{14761}{3800} & $V_{4}$ & Damsel & Action & \appbar{129537}{75905} & $W_{4}$ & twitter.com & Social media networks & \webbar{162500000000}{9600000000} \\
$A_{5}$ & YouTube VR & Entertainment & \appbar{14761}{3233} & $V_{5}$ & Avatar: The Last Airbender & Action & \appbar{129537}{65931} & $W_{5}$ & wikipedia.org & Dictionaries \& encyclopedias & \webbar{162500000000}{9200000000} \\
$A_{6}$ & TRIPP & Relaxation & \appbar{14761}{2898} & $V_{6}$ & Fool Me Once & Crime & \appbar{129537}{52113} & $W_{6}$ & instagram.com & Social media networks & \webbar{162500000000}{8000000000} \\
$A_{7}$ & Bait & Games & \appbar{14761}{2868} & $V_{7}$ & One Day & Comedy & \appbar{129537}{39797} & $W_{7}$ & reddit.com & Social media networks & \webbar{162500000000}{7400000000} \\
$A_{8}$ & Meta Quest Browser & Productivity & \appbar{14761}{2769} & $V_{8}$ & Lift & Action & \appbar{129537}{37603} & $W_{8}$ & amazon.com & eCommerce \& shopping & \webbar{162500000000}{4300000000} \\
$A_{9}$ & Blaston & Games & \appbar{14761}{2667} & $V_{9}$ & Griselda & Biography & \appbar{129537}{36651} & $W_{9}$ & duckduckgo.com & Search engines & \webbar{162500000000}{4200000000} \\
$A_{10}$ & Netflix & Media \& Streaming & \appbar{14761}{2579} & $V_{10}$ & Spaceman & Adventure & \appbar{129537}{30652} & $W_{10}$ & yahoo.com & Search engines & \webbar{162500000000}{4100000000} \\
$A_{11}$ & Gods of Gravity & Games & \appbar{14761}{1636} & $V_{11}$ & The Brothers Sun & Action & \appbar{129537}{21952} & $W_{11}$ & tiktok.com & Online TV & \webbar{162500000000}{3500000000} \\
$A_{12}$ & Cards \& Tankards & Games & \appbar{14761}{1374} & $V_{12}$ & Boy Swallows Universe & Crime & \appbar{129537}{16855} & $W_{12}$ & fandom.com & Dictionaries \& encyclopedias & \webbar{162500000000}{3200000000} \\
$A_{13}$ & Gravity Sketch & Creativity \& Design & \appbar{14761}{1073} & $V_{13}$ & Ripley & Crime & \appbar{129537}{14743} & $W_{13}$ & weather.com & Weather & \webbar{162500000000}{3100000000} \\
$A_{14}$ & Multiverse & Social & \appbar{14761}{979} & $V_{14}$ & Irish Wish & Comedy & \appbar{129537}{13188} & $W_{14}$ & whatsapp.com & Social media networks & \webbar{162500000000}{3100000000} \\
$A_{15}$ & HOLOFIT & Health \& Fitness & \appbar{14761}{946} & $V_{15}$ & Orion and the Dark & Animation & \appbar{129537}{12150} & $W_{15}$ & bing.com & Search engines & \webbar{162500000000}{3000000000} \\
$A_{16}$ & Half+Half & Games & \appbar{14761}{862} & $V_{16}$ & Code 8: Part II & Action & \appbar{129537}{11125} & $W_{16}$ & openai.com & Computer technology & \webbar{162500000000}{2600000000} \\
$A_{17}$ & Meta Quest Move & Utility & \appbar{14761}{805} & $V_{17}$ & The Greatest Night in Pop & Documentary & \appbar{129537}{10873} & $W_{17}$ & yandex.ru & Social media networks & \webbar{162500000000}{2500000000} \\
$A_{18}$ & VZFit & Sports & \appbar{14761}{601} & $V_{18}$ & Scoop & Biography & \appbar{129537}{9878} & $W_{18}$ & linkedin.com & Social media networks & \webbar{162500000000}{2000000000} \\
$A_{19}$ & Ultimechs & Games & \appbar{14761}{591} & $V_{19}$ & Badland Hunters & Action & \appbar{129537}{9058} & $W_{19}$ & microsoft.com & Computer technology & \webbar{162500000000}{1900000000} \\
$A_{20}$ & Liminal & Lifestyle & \appbar{14761}{543} & $V_{20}$ & The Abyss & Action & \appbar{129537}{6986} & $W_{20}$ & twitch.tv & Online TV & \webbar{162500000000}{1900000000} \\
$A_{21}$ & Spatial & Social, Utilities & \appbar{14761}{521} & $V_{21}$ & House of Ninjas & Action & \appbar{129537}{5750} & $W_{21}$ & live.com & Online TV & \webbar{162500000000}{1800000000} \\
$A_{22}$ & Neverboard & Games & \appbar{14761}{476} & $V_{22}$ & Parasyte: The Grey & Action & \appbar{129537}{5661} & $W_{22}$ & quora.com & Social media networks & \webbar{162500000000}{1700000000} \\
$A_{23}$ & Alcove & Social & \appbar{14761}{458} & $V_{23}$ & Baby Reindeer & Biography & \appbar{129537}{5647} & $W_{23}$ & netflix.com & Online TV & \webbar{162500000000}{1700000000} \\
$A_{24}$ & ForeVR & Games & \appbar{14761}{392} & $V_{24}$ & Amar Singh Chamkila & Biography & \appbar{129537}{5493} & $W_{24}$ & office.com & Computer technology & \webbar{162500000000}{1600000000} \\
$A_{25}$ & Maloka & Lifestyle & \appbar{14761}{390} & $V_{25}$ & The Signal & Drama & \appbar{129537}{5336} & $W_{25}$ & tsyndicate.com & Computer technology & \webbar{162500000000}{1500000000} \\
$A_{26}$ & Sphere Toon & Media \& Streaming & \appbar{14761}{266} & $V_{26}$ & Mea Culpa & Crime & \appbar{129537}{5190} & $W_{26}$ & bit.ly & Computer technology & \webbar{162500000000}{1500000000} \\
$A_{27}$ & Hoame & Health \& Fitness & \appbar{14761}{189} & $V_{27}$ & A Killer Paradox & Comedy & \appbar{129537}{4840} & $W_{27}$ & globo.com & News \& media publishers & \webbar{162500000000}{1400000000} \\
$A_{28}$ & MLB VR & Media \& Streaming & \appbar{14761}{186} & $V_{28}$ & The Tearsmith & Drama & \appbar{129537}{4333} & $W_{28}$ & imdb.com & Social media networks & \webbar{162500000000}{1300000000} \\
$A_{29}$ & Noda & Productivity & \appbar{14761}{166} & $V_{29}$ & The Beautiful Game & Drama & \appbar{129537}{3683} & $W_{29}$ & vk.com & Social media networks & \webbar{162500000000}{1200000000} \\
$A_{30}$ & Instagram & Social & \appbar{14761}{155} & $V_{30}$ & Supersex & Biography & \appbar{129537}{3068} & $W_{30}$ & cnn.com & News \& media publishers & \webbar{162500000000}{1200000000} \\
$A_{31}$ & vTime & Social & \appbar{14761}{140} & $V_{31}$ & Einstein and the Bomb & Documentary & \appbar{129537}{3063} & $W_{31}$ & manganato.com & News \& media publishers & \webbar{162500000000}{1100000000} \\
$A_{32}$ & Immerse & Lifestyle & \appbar{14761}{120} & $V_{32}$ & What Jennifer Did & Documentary & \appbar{129537}{2840} & $W_{32}$ & x.com & Social media networks & \webbar{162500000000}{1100000000} \\
$A_{33}$ & VR Workout & Health \& Fitness & \appbar{14761}{104} & $V_{33}$ & Testament: The Story of Moses & Documentary & \appbar{129537}{2562} & $W_{33}$ & pinterest.com & Social media networks & \webbar{162500000000}{1000000000} \\
$A_{34}$ & Facebook & Social & \appbar{14761}{103} & $V_{34}$ & Through My Window: Looking at You & Comedy & \appbar{129537}{2525} & $W_{34}$ & doubleclick.net & Business & \webbar{162500000000}{957100000} \\
$A_{35}$ & REMIO & Productivity & \appbar{14761}{87} & $V_{35}$ & My Name is Loh Kiwan & Drama & \appbar{129537}{2464} & $W_{35}$ & aliexpress.com & eCommerce \& shopping & \webbar{162500000000}{952900000} \\
$A_{36}$ & MeetinVR & Social & \appbar{14761}{78} & $V_{36}$ & The Wages of Fear & Action & \appbar{129537}{2390} & $W_{36}$ & ebay.com & eCommerce \& shopping & \webbar{162500000000}{872700000} \\
$A_{37}$ & Flipside Studio & Productivity & \appbar{14761}{72} & $V_{37}$ & Furies & Action & \appbar{129537}{2389} & $W_{37}$ & discord.com & Social media networks & \webbar{162500000000}{864100000} \\
$A_{38}$ & Prisms MATH & Social & \appbar{14761}{64} & $V_{38}$ & Delicious in Dungeon & Animation & \appbar{129537}{2254} & $W_{38}$ & sharepoint.com & Business & \webbar{162500000000}{842600000} \\
$A_{39}$ & Zoe & Social & \appbar{14761}{60} & $V_{39}$ & Turning Point: The Bomb \& Cold War & Documentary & \appbar{129537}{2066} & $W_{39}$ & zoom.us & Computer technology & \webbar{162500000000}{838100000} \\
$A_{40}$ & Arthur & Social & \appbar{14761}{59} & $V_{40}$ & Bandidos & Action & \appbar{129537}{1921} & $W_{40}$ & spotify.com & Music & \webbar{162500000000}{835700000} \\
$A_{41}$ & VirtualSpeech & Productivity & \appbar{14761}{41} & $V_{41}$ & Shirley & Biography & \appbar{129537}{1831} & $W_{41}$ & indeed.com & Business & \webbar{162500000000}{833600000} \\
$A_{42}$ & Pluto TV & Media \& Streaming & \appbar{14761}{39} & $V_{42}$ & Crooks & Action & \appbar{129537}{1173} & $W_{42}$ & github.com & Computer technology & \webbar{162500000000}{831000000} \\
$A_{43}$ & Messenger & Utilities & \appbar{14761}{21} & $V_{43}$ & Heart of the Hunter & Action & \appbar{129537}{1117} & $W_{43}$ & msn.com & Social media networks & \webbar{162500000000}{803100000} \\
$A_{44}$ & Smartsheet & Productivity & \appbar{14761}{19} & $V_{44}$ & Love, Divided & Comedy & \appbar{129537}{1063} & $W_{44}$ & canva.com & Business & \webbar{162500000000}{794000000} \\
$A_{45}$ & Nanome & Productivity & \appbar{14761}{14} & $V_{45}$ & The Antisocial Network & Documentary & \appbar{129537}{1026} & $W_{45}$ & bbc.com & News \& media publishers & \webbar{162500000000}{791000000} \\
$A_{46}$ & Resolve & Productivity & \appbar{14761}{11} & $V_{46}$ & Woody Woodpecker Goes to Camp & Adventure & \appbar{129537}{662} & $W_{46}$ & foxnews.com & News \& media publishers & \webbar{162500000000}{748500000} \\
$A_{47}$ & Softspace & Productivity & \appbar{14761}{8} & $V_{47}$ & The Hijacking of Flight 601 & Drama & \appbar{129537}{580} & $W_{47}$ & taboola.com & Business & \webbar{162500000000}{720000000} \\
$A_{48}$ & Monday.com & Productivity & \appbar{14761}{5} & $V_{48}$ & The Tourist & Action & \appbar{129537}{454} & $W_{48}$ & espn.com & News \& media publishers & \webbar{162500000000}{705900000} \\
$A_{49}$ & LastPass & Utilities & \appbar{14761}{5} & $V_{49}$ & Nuevo rico, nuevo pobre & Comedy & \appbar{129537}{262} & $W_{49}$ & paypal.com & Computer technology & \webbar{162500000000}{703900000} \\
$A_{50}$ & STARZ & Media \& Streaming & \appbar{14761}{2} & $V_{50}$ & The Cartel: The Origin & Drama & \appbar{129537}{254} & $W_{50}$ & samsung.com & Computer technology & \webbar{162500000000}{684200000} \\
\bottomrule
\end{tabular}%
}
\end{table}

\clearpage
\twocolumn

\newpage 


\section{Meta-Review}

The following meta-review was prepared by the program committee for the 2026 IEEE Symposium on Security and Privacy (S\&P) as part of the review process as detailed in the call for papers.

\subsection{Summary}
The paper presents BRAVESPY, a VR side-channel attack that uses unrestricted motion sensors in commercial VR headsets to reconstruct the user's brain-wave signal and extract privacy-sensitive information. BRAVESPY leverages subtle pupillary-response-induced vibrations captured through motion sensors and translates them into EEG signals to infer user activity and cognition.

\subsection{Scientific Contributions}
\begin{itemize}
\item 5. Identifies an impactful vulnerability.
\item 6. Provides a valuable step forward in an established field.
\end{itemize}

\subsection{Reasons for Acceptance}
\begin{enumerate}
\item The paper identifies an impactful privacy risk in VR systems. BRAVESPY showed that unrestricted motion sensors can be translated into brain-induced EEG signals that expose substantially more privacy information.
\item The paper provides a valuable step forward in an established field. The paper goes beyond prior VR motion-sensor privacy work by attempting to connect subtle headset vibrations with higher-level perceptual inference. The experimental effort is substantial, involving multiple devices, multiple users, and several inference tasks.
\item The paper presents a broad effectiveness and accuracy evaluation. The evaluation spans multiple privacy-inference tasks across UI-, user-, and brain-level settings and covers multiple VR headsets, users, sampling rates, and environmental conditions. The reported results indicate strong performance in several scenarios, especially for UI-level inference, while also showing measurable accuracy for the more challenging brain-level task.
\end{enumerate}

\subsection{Noteworthy Concerns} 
None

\end{document}